\documentclass{aa} % for the abstract without structuration 
\usepackage{graphicx}
\usepackage{txfonts}
\usepackage{epstopdf}

\begin{document}
   \title{Multi-zone warm and cold clumpy absorbers in 3 Seyfert galaxies}

   \subtitle{}

   \author{C. Ricci
          \inst{1,2},
          V. Beckmann
          \inst{3},
           M. Audard
           \inst{1,2}
            \and
           T. J.-L. Courvoisier\inst{1,2}
          }

   \institute{ \textsl{ISDC} Data Centre for Astrophysics, University of Geneva, ch. d'Ecogia 16, 1290 Versoix, Switzerland 
    \and Geneva Observatory, University of Geneva, ch. des Maillettes 51, 1290 Versoix, Switzerland 
    \and APC, Fran\c{c}ois Arago Centre, Universit\'e Paris Diderot, CNRS/IN2P3, 10 rue A. Domon et L. Duquet, 75205 Paris Cedex, France\\
             }
    \offprints{e-mail: Claudio.Ricci@unige.ch} 
   \authorrunning{ C. Ricci et al.}
   \titlerunning{Multi-zone warm and cold clumpy absorbers in 3 Seyfert galaxies}
    \date{Received; accepted}

% \abstract{}{}{}{}{} 
% 5 {} token are mandatory
 
 \abstract
{}
  {We present the first detailed X-ray analysis of three AGN, the Seyfert~1 galaxies UGC~3142 and ESO~140-43, and the Seyfert~2 galaxy ESO~383-18, in order to study the geometry and the physical characteristics of their absorbers.}
  {High quality XMM-Newton EPIC and RGS data were analysed, as well as Swift/XRT and BAT and INTEGRAL IBIS/ISGRI data, in order to cover the 0.3--110  keV energy range. For ESO~140-43 also XMM-Newton/OM and Swift/UVOT data were used. We studied the variability of the three AGN on a time-scale of seconds using the EPIC/PN light curves, and the long-term time-scale variability of ESO~140-43 using two observations performed 6 months apart by XMM-Newton.}
  {The spectra of the three Seyfert galaxies present a "soft excess'' at energies E~$ < 2 \rm \,keV$ above a power-law continuum that can be modeled by complex absorption, without any additional emission component. The X-ray sources in UGC~3142 and ESO~383-18 are absorbed by two layers of neutral material, with covering fractions $\rm \,f_1 \simeq 0.92$ and $\rm \,f_2 \simeq 0.57$ for UGC~3142, and $\rm  \,f_{1} \simeq 0.97$ and $\rm  \,f_{2} \simeq 0.86$ for ESO~383-18. While the clumpy absorber could be part of a disc wind or of the broad line region for UGC~3142, in the case of ESO~383-18 a clumpy torus plus Compton thin dust lanes are more likely. The spectra of ESO~140-43 can be well fitted using a power law absorbed by three clumpy ionized absorbers with different covering factors, column densities, and ionization parameters, likely part of a moving clumpy system, which might be a disc wind or the broad line region.
The strong spectral and flux variability on a time scale of 6 months seen in ESO~140-43 is likely due to changes in the moving absorbers. The variation of the covering factor of one of the three ionized absorbers could be detected, on a kilo-seconds time-scale, in the EPIC light-curve of ESO~140-43. }
    {}
%}


   \maketitle
   %______________________________________________________________
   
\section{Introduction}
Active Galactic Nuclei (AGN) are thought to be powered by accretion onto super massive black holes. In this picture the X-ray emission is produced by UV photons from the innermost edge of the accretion disc, which undergo multiple inverse Compton scatterings by a population of hot electrons located in a coronal region sandwiching the disc (Haardt \& Maraschi 1991, 1993). The spectrum in the X-rays can be well approximated by a power law with an exponential cut-off at energy $E_C\simeq kT_e$, where $T_e$ is the temperature of the electrons. Besides the power-law like continuum, two prominent features generated by the reflection of the continuum are observed in the X-ray spectra of AGN: a neutral iron K$\alpha$ line (at 6.4 keV in the  local reference frame) and a reflection hump which peaks at E~$\simeq 30$ keV (Magdziarz \& Zdziarski 1995). Not all these components are clearly observed in all AGN (e.g., Soldi et al. 2005, Beckmann et al. 2004).

An alternative model for the accretion process onto black holes is that of clumpy accretion flows (e.g., Guilbert \& Rees 1988). Courvoisier \& T\"urler (2005) assume that the different elements (clumps) of the accretion flow have velocities that may differ substantially. As a consequence, collisions between these clumps will appear when the clumps are close to the central object, resulting in emission  (Ishibashi \& Courvoisier 2009).

At optical energies, AGN are classified according to their emission lines into type 1 (showing broad and narrow emission lines), and type 2 (showing only narrow lines). A similar distinction between the two types can be done in the X-rays. Here, type 2 objects show absorption which supresses the soft X-ray spectrum. The dividing line between unabsorbed and absorbed AGN is often set at a hydrogen column density of $N_{\rm H} = 10^{22} \rm \, cm^{-2}$. X-ray data show that most AGN unabsorbed in X-ray are optical Seyfert~1 type, and most, but not all, AGN that are absorbed belong to the Seyfert~2 group (e.g., Awaki et al. 1991).
Antonucci (1993) proposed that the extinguishing region forms a torus around the central region. The presence of an absorbing torus guarantees anisotropic obscuration of the central region, so that the sources viewed face-on are type 1, and those observed edge-on are type 2 (for which, although it is highly extinguished, the broad line region exists). AGN spectra are mainly affected by absorption at energies $\lesssim 2$ keV, with the effect depending on whether the absorber is cold (neutral or weakly ionized), or warm (highly ionized). The geometry of the absorber also plays a major role. When it does not cover completely the source (partially covering absorber), some continuum leaks out and reaches the observer, resulting in an apparent excess over the continuum at higher energies. The {\it Einstein} satellite provided the first evidence that the absorbing gas can cover only a fraction of the line of sight towards the AGN (Reichert et al. 1985, Holt et al. 1980).
The partial covering scenario can also explain some of the observed X-ray variability, which could be due to rearrangements of the cloud distribution (e.g., Abrassart \& Czerny 2000).

The first evidence of absorption due to warm material came from studies of the {\it Einstein} X-ray spectrum of the QSO MR 2251-178 (Halpern 1984, Pan et al. 1990), and from AGN spectra taken with {\it ROSAT} (Turner et al. 1993). Since then, highly ionized absorbers have been observed in about half of the X-ray spectra of type 1 AGN, both Seyfert 1 (e.g., Reynolds 1997, George et al. 1998) and quasars (Piconcelli et al. 2005), with column densities up to $\gtrsim 10^{23} \rm \,cm^{-2}$, and often consisting of several zones of ionized gas (e.g., Nandra et al. 1993, Pounds et al. 1994, Kaspi et al. 2002, Steenbrugge et al. 2005).
In the case of NGC~3783, Netzer et al. (2003) and Krongold et al. (2003) discussed the observed complexity in the context of several phases of absorbing gas having distinct temperatures and ionization states. Evidence of a multi-phase warm absorber was also recently reported by Longinotti et al. (2010). The observed absorption lines are often blue-shifted with respect to the optical emission lines, which implies that they are produced in an outflowing region, with mean velocities covering the range from hundreds to thousands of $\rm \,km \,s^{-1}$ (Kaspi et al. 2002, Krongold et al. 2003, Blustin et al. 2005, McKernan et al. 2007, Blustin et al. 2007). Analyzing NGC~3516, Turner et al. (2005) discovered a high column absorber with a hydrogen column density of $N_{\rm \,H}\sim 10^{23} cm^{-2}$ and a ionization parameter of $\log \xi \sim 2$, covering about 50\% of the continuum source. Changes in the covering fraction of the absorber could explain the spectral variability observed in this source (Turner et al. 2008).
A complete review on the recent developments of absorption models can be found in Turner \& Miller (2009).

A soft ($E \lesssim 2$ keV) excess over the power-law component dominant at higher energies has been found in the X-ray spectra of many Seyfert galaxies (Saxton et al. 1993). The origin of the soft excess is still an open issue. In the past the soft excess was often associated with the high-energy tail of the thermal emission of the disc, but it has been recently shown that the temperature of the disc should be constant (0.1--0.2 keV), regardless of the mass and luminosity of the AGN (Gierlinski \& Done 2004). This result implies that some other mechanism is at work, as the temperature of the disc should depend on both the mass of the black hole and the accretion rate.
Three competing models have been brought forward in order to explain the soft X-ray excess: i) an additional Comptonization component (e.g., Dewangan et al. 2007), ii) ionized reflection (e.g. from the disk, e.g. Crummy et al. 2006), iii) complex and/or ionized absorption (e.g., Gierlinski \& Done 2004).
Chevallier et al. (2006) examined these three main possibilities and showed that, although the absorption model can successfully fit the data of several AGN, for many AGN the quality of the data does not allow a firm conclusion. Nevertheless, they favour a reflection model absorbed by a modest relativistic wind.
Done \& Nayakshin (2007) agree that the excess is most probably due to partially ionized material moving at relativistic speeds close to the black hole. The two potential scenarios for this effect would be: i) an accretion disk seen in reflection, ii) a wind above the disk, seen in absorption. They show that the reflection model would require very specific ionization parameters, contrary to the disk wind model.
Schurch \& Done (2008) pointed out that, although the absorption model using partially ionized material can be physically interpreted as a radiatively driven accretion disc wind, these winds would require velocities of $\sim 0.9 \rm c$ in order to reproduce the soft X-ray excess. They conclude that if the soft excess is produced by absorption, it seems more likely that the material is clumpy and/or only partially covers the source rather than forming a continuous outflow.

The three AGN discussed here are part of the {\it INTEGRAL} AGN catalogue (Beckmann et al. 2009). As at the time of compilation of the catalogue no absorption information was available for a number of sources, unpublished {\it XMM-Newton} and {\it Swift}/XRT data were used to constrain $N_{\rm H}$. The three Seyfert galaxies UGC 3142, ESO 140-43, and ESO 383-18 are those sources which showed particularly strong evidence for excess in the soft X-ray spectra which requires more complex modeling of the absorption.

UGC~3142 is a nearby Seyfert~1 galaxy and was first detected in X-ray by {\it ROSAT} (Boller et al. 1992), and in hard X-ray by {\it INTEGRAL} (Bassani et al. 2006).
ESO~383-18 is a Seyfert 2 galaxy, detected in hard X-ray by {\it INTEGRAL} (Krivonos et al. 2007).
ESO~140-43, also known as Fairall~51, is a Seyfert 1. The first X-ray detection of this AGN was done by {\it HEAO A-2} (Marshall et al. 1979), and subsequently it was detected by {\it ROSAT} (Boller et al. 1992). At hard X-rays ESO~140-43 was detected by {\it INTEGRAL} IBIS/ISGRI (Beckmann et al. 2007). Information on the 3 sources are reported in Table \ref{info_sources}.
We present here the first detailed X-ray spectral analysis of these three Seyfert AGN using {\it XMM-Newton}, {\it Swift} and {\it INTEGRAL} data. The data analysis is described in Section \ref{data_analysis}. In Section \ref{spectral_analysis}, we discuss the spectral modelling of the X-ray spectra, and we present the study of the X-ray and optical/UV variability in Section \ref{X_UV_variability}. In Section \ref{discussion}, we discuss the results obtained, and we present the conclusions in Section \ref{conclusions}.

\begin{table*}
\caption[]{Positions (J2000), redshift and Galactic hydrogen column densities (from Dickey \& Lockman 1990) of the 3 sources analyzed in this work.}
\label{info_sources}
\begin{center}
\begin{tabular}{llllll}
\hline
Object & Type &RA &  DEC & z & Galactic N$_{\rm \,H}$\\
& & & & &{\scriptsize [$10^{21}\rm \,cm^{-2}$]} \\
\hline
UGC~3142  & Seyfert 1 &$\rm 04^{\,h}43^{\,m}46.8^{\,s}$ & $+28^{\circ }58'19"$ & 0.0217 & 1.9 \\
ESO~383-18  & Seyfert 2 &$\rm 13^{\,h}33^{\,m}26.3^{\,s}$ & $-34^{\circ }00'59"$ & 0.0124 & 0.42\\
ESO~140-43  & Seyfert 1 &$\rm 18^{\,h}44^{\,m}54.0^{\,s}$ & $-62^{\circ }21'53"$ & 0.0141 & 0.73\\
\hline

\hline
\end{tabular}
\end{center}
\end{table*}
\section{Data analysis}\label{data_analysis}
The journal of the observations is shown in Table~\ref{obslog}.

\subsection{{\it XMM-Newton} EPIC, RGS and OM}\label{epic_analysis}
UGC~3142 was observed by the European Photon Imaging Camera (EPIC) on board of the  {\it XMM-Newton} space observatory (Jansen et al. 2001) in March 2007. The EPIC observation was performed using the PN camera (Struder et al. 2001) and the MOS (1 and 2) cameras (Turner et al. 2001), all of them operating in Prime Full Window mode, and all equipped with the ``Medium" blocking filter.

ESO~383-18 was observed by {\it XMM-Newton} in October 2006 with the three EPIC cameras operating in Prime Full Window mode. PN and MOS~1 cameras were used with the ``Thin" blocking filter, while MOS~2 was using the ``Medium" filter.

ESO~140-43 was observed twice by {\it XMM-Newton}. The first observation  was performed in September 2005 (henceforth 2005-observation), while the second observation  was carried out in March 2006 (henceforth 2006-observation). In both observations the three EPIC cameras were equipped with the``Medium" blocking filter and operated in Prime Full Mode.

We checked the background light curve in the 10--12 keV energy band for PN and above 10 keV for MOS in order to detect and filter the exposures for periods of high background activity. The recommended thresholds for the background count rate %of these light curves 
are 0.35  ct/s and 0.4  ct/s for the MOS and PN camera, respectively. For UGC~3142, only the PN is strongly affected, and the exposure is reduced from 10 ks to 8.5 ks. In the case of ESO~383-18 this does not change anything in the good time intervals (GTI), as the background count rates of the observations are well below the recommended thresholds.
Particular care has to be taken in the case of ESO~140-43, as the EPIC detectors experienced a strong flux of high energy particles during the observations. During the 2005-observation this flux was between 0.4 and 2.2 ct/s for the PN, while it was much below the recommended threshold for MOS.
The contribution of the background to the spectrum of the 2005-observation is weak ($\ll$ 1\%) over the range considered in the analysis (0.3--10 keV), thus we filtered only events with a background count rate higher than 1 ct/s, which reduces the  GTI to 22.5 ks, excluding the last 2.5 ks of the observation. The 2006-observation presented a stronger flux of high-energy particles, with a maximum flux of 5 ct/s in the case of PN and of 1.5 ct/s in the case of MOS. Also in this observation the contribution of the background is not strong ($\simeq 8 \%$), and we set the background count rate threshold at 4 and 0.8 ct/s for the PN and MOS, respectively. This reduces the exposure time to 18 ks for the PN, and to 17.6 ks for the two MOS cameras.

We checked for the presence of pile-up using the {\it epatplot} task, and we found that it is negligible in all observations. For the analysis of EPIC data we used the {\it XMM-Newton} Standard Analysis Software (SAS) version 8.0.0, which also generates the ancillary response matrices (ARFs) and the detector response matrices (RMFs), through the tasks {\it arfgen} and {\it rmfgen}. We used the {\it epchain} and {\it emchain} tasks for processing the raw PN and MOS data files, respectively. We selected only patterns corresponding to single, double, triple and quadruple events for EPIC/MOS (PATTERN~$\leq 12$), while for EPIC/PN we considered only single and double events (PATTERN~$\leq 4$). The source spectra were extracted from the final filtered event list using circular regions centered on the sources, while the background was estimated from regions close to the source (on the same CCD), where no source was present. The same regions were used to estimate the PN light curves of the sources, which were corrected for the influence of the background using the task {\it epiclccorr}.  The radii of these regions are listed in Table~\ref{radii}. The spectra obtained were grouped to have at least 20 counts per bin, in order to use $\chi^{2}$ statistics.

The RGS spectra were extracted using the {\it rgsproc} task, and only the first-order spectra of the two RGS detectors were analyzed. The spectra were rebinned by a factor of 16.\newline

In order to have information about the flux of the {3 sources} in the optical and ultraviolet, we also analyzed the Optical/UV Monitor telescope (OM, Mason et al. 2001) data using the task {\it omichain}.

\subsection{{\it Swift} XRT, BAT and UVOT}\label{swift_analysis}
Two pointed {\it Swift} observations of ESO~140-43 have also been analysed. {\it Swift} (Gehrels et al. 2004) carries two narrow-field instruments, XRT (Burrows et al. 2005) with energy range 0.3--10.0 keV and UVOT (Roming et al. 2005) which is equipped with 6 filters in the optical and UV range. The XRT and UVOT data have been analysed using HEADAS version 6.6.1, and applying the latest calibration files as available in January 2009. The analysis of XRT data was performed with the extraction of the source and the background spectra from the event lists using circular regions, in a similar way to what has been done for the EPIC analysis. The ancillary response matrices have been generated using the task {\it xrtmkarf}. The spectra have been grouped in order to have at least 20 counts per bin.\newline 
ESO~140-43 and UGC~3142 have been detected by {\it Swift}/BAT in the 22 month survey (Tueller et al. 2010) in the 14--195 keV band with a significance of $4.6\sigma$ and $4.9\sigma$, respectively. The spectra used in this work are provided by the BAT team\footnote{http://swift.gsfc.nasa.gov/docs/swift/results/bs22mon/}.
ESO~383-18 has been detected with a significance of $8.6\sigma$ by BAT within the 39 month survey presented by Cusumano et al. (2009).\newline
The UVOT analysis was performed creating a source and a background region file, and then using the task {\it uvotsource} to extract the flux values.

\subsection{INTEGRAL IBIS/ISGRI}\label{isgri_analysis}
For the analysis of the IBIS/ISGRI spectra we used the observations for which the sources were at less than $10^{\circ}$ from the instrument axis. 
Note that because of the nature of coded mask imaging, the whole sky image taken by the instrument has to be considered in the analysis, as all sources in the field of view contribute to the background (Caroli et al. 1987). The exposures in Table~\ref{obslog} are the ISGRI effective on-source times. The analysis was performed using version 7.0 of the Offline Science Analysis (OSA) software distributed by the ISDC (Courvoisier et al. 2003).
The analysis of the IBIS/ISGRI data is based on a cross-correlation procedure between the recorded image on the detector plane and a decoding array derived from the mask pattern. The ISGRI spectra have been extracted using the standard method as described in Goldwurm et al. (2003): fluxes and count spectra are extracted at the source positions for each pointing in predefined energy bins. Spectra of the same source collected in different pointings are summed to obtain an averaged spectrum during the observation.

As the full width at half maximum (FWHM) angular resolution of IBIS/ISGRI is 12 arc-minutes we investigated for possible contamination from sources close to the 3 AGN using {\it Simbad}\footnote{http://simbad.u-strasbg.fr/} and {\it NED}\footnote{http://nedwww.ipac.caltech.edu/} and did not find any significative.

\begin{table*}
\caption[]{X-ray observation log.}
\label{obslog}
\begin{center}
\begin{tabular}{lllllll}
\hline
Object & Mission & Instrument &  Start time & End Time & Exposure & Observation ID \\
  &  &      &  \scriptsize{[U.T.]} &  \scriptsize{[U.T.]} &   \scriptsize{[ksec]}   &   \\
\hline
UGC~3142 & \textit{INTEGRAL} & IBIS/ISGRI&  February 2003 & October 2008 &  485 &-- \\
& \textit{XMM-Newton} & EPIC/PN & 2007-03-18 21:28:29  & 2007-03-19 00:15:51 &  8.5 & 0401790101  \\
& \textit{XMM-Newton} & EPIC/MOS & 2007-03-18 21:06:10 &  2007-03-19 00:19:51 &  11.6 & 0401790101  \\
& XMM-Newton & RGS & 2007-03-18  21:05:31 & 2007-03-19 00:23:59 &  11.9 & 0401790101\\
ESO~383-18 & \textit{INTEGRAL} & IBIS/ISGRI &  March 2003 & August 2007   & 260 &--\\
& \textit{XMM-Newton} & EPIC/PN & 2006-01-10 02:49:18 & 2006-01-10 06:50:06   & 14.5 & 0307000901\\
& \textit{XMM-Newton} & EPIC/MOS & 2006-01-10 02:26:59  & 2006-01-10 06:55:28   & 16.1& 0307000901\\
& XMM-Newton & RGS &  2006-01-10 02:26:21 & 2006-01-10 06:54:50  & 16.1 & 0307000901\\
ESO~140-43 & \textit{INTEGRAL}& IBIS/ISGRI& 2007-10-29 & 2007-04-11  & 379 &-- \\
& \textit{XMM-Newton} & EPIC/PN & 2005-09-08 03:09:46 & 2005-09-08 10:07:29  & 22.5& 0300240401\\
& \textit{XMM-Newton}& EPIC/MOS & 2005-09-08 02:47:28 & 2005-09-08 05:30:11  & 9.8& 0300240401\\
& XMM-Newton & RGS &  2005-09-08 02:46:44 & 2005-09-08 10:15:22 & 26.9  &0300240401\\
& \textit{XMM-Newton} & EPIC/PN & 2006-03-07 12:36:40 & 2006-03-07 18:19:46  & 18 & 0300240901\\
& \textit{XMM-Newton} & EPIC/MOS & 2006-03-07 12:14:22 & 2006-03-07 18:23:52  & 17.6 & 0300240901\\
& XMM-Newton & RGS & 2006-03-07 12:13:44 & 2006-03-07 18:27:15 & 22.4 & 0300240901\\
& \textit{Swift} & XRT & 2008-05-20 12:42:01 & 2008-05-20 14:25:13& 6.2 & 00037809001\\
& \textit{Swift} & XRT & 2008-05-25 11:33:00 & 2008-05-25 12:26:20 & 3.2&00037809002\\
\hline
\end{tabular}
\end{center}
\end{table*}

\begin{table*}
\caption[]{Background and source radii used for the extraction of {\it XMM-Newton}/EPIC spectra. The index 1 refers to the 2005 observation of ESO~140-43, while 2 to the 2006 observation.}
\label{radii}
\begin{center}
\begin{tabular}{lllllll}
\hline
Object & Instrument &  Source & Background & Instrument &  Source & Background \\
\hline
UGC~3142  & EPIC/PN & 34.5'' & 48.1" & EPIC/MOS & 32.2" & 62" \\
ESO~383-18  & EPIC/PN & 37" & 43.6" & EPIC/MOS & 24.2" & 47.8" \\
ESO~140-43$^1$  & EPIC/PN & 36.7" & 52.9" & EPIC/MOS & 37.5" & 66" \\
ESO~140-43$^2$  & EPIC/PN & 34" & 61.7" & EPIC/MOS & 26.5" & 50" \\
\hline

\hline
\end{tabular}
\end{center}
\end{table*}

\section{X-ray spectral analysis}\label{spectral_analysis}
%%%%%%%%%%%%%%%%%%%%%%%%%%%%%%%%%%%%%%%%%%%%%%%%%%%%%%%%%%%%%%%%%%%%%%%%%%%%%%%%%%%%%%
The spectra derived from {\it INTEGRAL}, {\it XMM-Newton} and {\it Swift} data were analyzed using XSPEC version 11.3.2 (Arnaud 1996). 
Throughout the paper we give symmetric and asymmetric errors which represent 1$\sigma$ and 3$\sigma$ confidence intervals, respectively. The probabilities that additive parameters improve a fit are calculated using the F-test. The errors on the fluxes, luminosities and the equivalent widths have been calculated at the 3$\sigma$ level, using 1000 Monte Carlo simulations.
In all cases, we added to the models a multiplicative factor to account for cross-calibration between the two EPIC detectors. We fixed the factor to 1 for EPIC/PN and left the factor for the two MOS detectors free to vary. The value of this factor turned out, as expected, to be close to 1 within a few percent (the differences in normalization of the two detectors should be below 5-10\%)\footnote{http://xmm2.esac.esa.int/docs/documents/CAL-TN-0052-5-0.eps.gz}.\newline 
The energy of the lines is calculated in the rest frame of the host galaxies, including in the model their redshifts.
\subsection{UGC~3142}
\begin{figure}[t!]
\centering
\includegraphics[width=9cm]{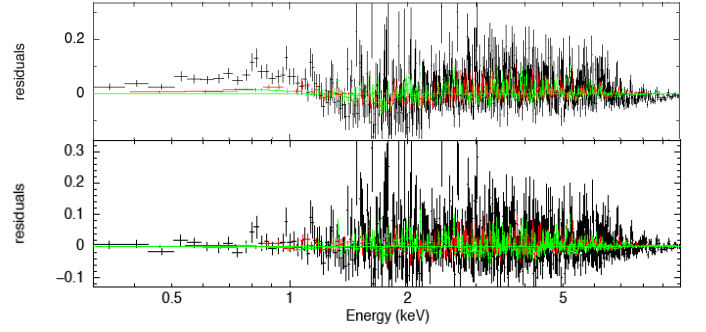}
\caption{UGC~3142 EPIC/PN (in black) and MOS residuals obtained with the simple power law model (upper panel) and the best (lower panel) model, respectively, as described in the first and third row of Table \ref{UGC_res_tab}.}
\label{fig:UGC_res}
\end{figure}

The 0.3--10 keV EPIC/PN and MOS  spectrum of UGC~3142 cannot be well represented using a simple power law model absorbed by cold matter, resulting in $\chi^{2}$=1488 for 1028 degrees of freedom (dof). The absorbed power law gives a very flat photon index ($\Gamma \simeq 0.9$) and results in a strong excess (Fig. \ref{fig:UGC_res}) in soft X-ray ($E\lesssim1$ keV). 
In order to account for the excess we tested several models, taking into account different emission processes. Considering a thermal origin of the excess, we added an accretion disc model consisting of multiple blackbody components (implemented as {\it diskbb} in XSPEC) to the power law. Such a model does not yield a good fit ($\chi^2 = 1225$ for 1026 dof), and the residuals still show a strong excess below 1 keV. We then modeled the excess with a second Comptonization region, using the {\it compTT} model of Titarchuk (1994). This second Comptonization region could be identified as a hot disc surface layer. Adding this model to the power-law, and considering absorption, yields a good fit ($\chi^2 = 1035$ for 1024 dof), but with an unlikely very steep photon index of $\Gamma = 6.4^{+1.4}_{-1.9}$, and an overly strong Compton component. As an additional emission component cannot model the soft excess, we tested whether absorption could explain the observed excess. % using a partially covering absorber.
We used a multiplicative component for the absorption in order to account for a neutral absorber partially covering the X-ray source (implemented as {\it pcfabs} in XSPEC):
\begin{equation}
\label{eq_pcfabs}
M(E)=fe^{-N_{\rm \,H}\sigma(E)}+(1-f),
\end{equation}
where the parameter $f$ is the covering fraction, which varies between 0 (absorber out of the line of sight) and 1 (absorber completely covering the source).

Using a power-law model absorbed by Galactic cold matter plus a layer of partially absorbing neutral material improves significantly the results at the 99.9\% confidence level ($\chi^2 = 1080$ for 1027 dof), and steepens the power law ($\Gamma=1.13^{+0.04}_{-0.05}$). 
This power law is still quite flat, and the model does not connect with the hard X-ray spectrum obtained by {\it INTEGRAL} IBIS/ISGRI and {\it Swift}/BAT. Adding another layer of partially absorbing neutral material improves the fit again at the 99.9\% confidence level ($\chi^2 = 1030$ for 1025 dof), and results in a photon index of $\Gamma=1.4^{+0.1}_{-0.1}$. The two partially covering absorbers present different characteristics: one of them covers 92\% of the X-ray emitting source and has a hydrogen column density of $N_{\rm H}=1.4_{-0.2}^{+0.2} \times 10^{22}\rm \,cm^{-2}$, while the other has a smaller covering factor ($\simeq 57\%$) and a higher column density ($N_{\rm H}=5_{-1}^{+2} \times 10^{22}\rm \,cm^{-2}$).

Although the model fits well the data, an excess is still present around 6.4 keV. Adding a Gaussian line improves the fit at the 99.9\% confidence level ($\chi^2 = 994$ for 1022 dof). The line is located at $6.38^{+0.03}_{-0.03}$ keV, being thus consistent with the iron K$\alpha$ line, and has a width of $\sigma=41^{+32}_{-29}$ eV, and an equivalent width of EW=$83^{+189}_{-60}$ eV. The X-ray spectra of AGN can show the presence of an additive broad line at 6.7 keV due to ionized iron (e.g., Matt et al. 2001). We investigated the presence of this feature adding another Gaussian line to our best model, and fixing the energy of the line at 6.7 keV, but this did not improve significantly the result ($\Delta \chi ^2 \simeq 0$).

The RGS spectrum of UGC~3142 has a very low signal to noise ratio, and has not been analyzed.

In order to search for the presence of a cut-off or a reflection hump in hard X-ray, we combined the EPIC data with the {\it Swift}/BAT and {\it INTEGRAL} IBIS/ISGRI spectra. The 0.3--110 keV spectrum is well represented ($\chi ^2$ =1011 for 1034 dof) by the best model for the EPIC data alone, as presented above (power-law absorbed by two layers of partially covering material). Adding a high-energy cut-off does not significantly improve the fit ($\chi ^2$ =1008 for 1033 dof). The high-energy cut-off is nevertheless constrained ($E_C=63^{+118}_{-23}$ keV), but higher quality hard X-ray data will be necessary to verify it. As the observations were not simultaneous we added a cross-calibration constant to the model, which resulted to be $C_I=1.4$ and $C_B=0.5$, for ISGRI and BAT, respectively.
We also tested the presence of a Compton reflection hump using the {\it pexrav} model in XSPEC, which implies the presence of reflection due to neutral material. We fixed the inclination angle at $i=30^{\circ}$ (as a common assumption for type 1 AGN, see e.g. Matt et al. 2006, Decarli et al. 2008), and found that the data are consistent with such a feature, although the reflection normalization $R$ and the cut-off energy $E_C$ are not constrained, and have a strong degeneracy. In particular the value of $R$ is consistent with no reflection (R=$1^{+2}_{-1}$). 

We can thus conclude that the best model for the 0.3--110 keV spectrum of UGC~3142 is a power-law absorbed by two layers of partially covering material plus a Gaussian line. The results of the spectral analysis are reported in Table \ref{UGC_res_tab}.

\begin{table*}
\caption[]{Results obtained from the X-ray spectral analysis of UGC~3142. The data have been modeled using an absorbed power law model ($a$), a power law absorbed by Galactic cold matter plus a layer of partially absorbing neutral material ($b$), and a power law absorbed by Galactic cold matter and two layers of partially absorbing neutral material, plus  a Gaussian K$\alpha$ iron line ($c$). EPIC includes PN, MOS1 and MOS2 data.}
\label{UGC_res_tab}	
\begin{tabular}{lllllllll}
\hline
 Instruments & Energy&  $\chi ^2$ (dof)& $\Gamma$& $N_{\rm H}$ &$N_{\rm H}^1$&  f$^1$ & $N_{\rm H}^2$ &  f$^2$  \\
 & {\scriptsize [keV]} & &  & {\scriptsize [$10^{20}\rm \,cm^{-2}$]} & {\scriptsize [$10^{22}\rm \,cm^{-2}$]} & & {\scriptsize[$10^{22}\rm \,cm^{-2}$]} & \\
\hline
EPIC$^a$& 0.3--10& 1488 (1028)& $0.9^{+0.1}_{-0.1}$ & $1.4^{+0.1}_{-0.1}$ &  -- & -- &  -- &  -- \\
EPIC$^b$& 0.3--10& 1080 (1027)& $1.13^{+0.04}_{-0.05}$ & $0.19$ &  $2.1^{+0.1}_{-0.1}$ & $0.92^{+0.01}_{-0.01}$ & -- &  -- \\
EPIC$^c$& 0.3--10& 994 (1022)& $1.4^{+0.1}_{-0.1}$ & $0.19$ & $1.4^{+0.2}_{-0.2}$ & $0.92^{+0.01}_{-0.02}$ &  $5^{+1}_{-1}$ & $0.57^{+0.08}_{-0.09}$ \\
(EPIC + ISGRI + BAT)$^c$ & 0.3--110 & 1011 (1034) & $1.5^{+0.1}_{-0.1}$ & $0.19$ & $1.4^{+0.3}_{-0.4}$ & $0.92^{+0.02}_{-0.06}$ &  $4^{+2}_{-1}$ & $0.6^{+0.2}_{-0.1}$  \\
\hline
\end{tabular}
\end{table*}
\begin{figure}[t!]
\centering
\includegraphics[height=9cm,angle=270]{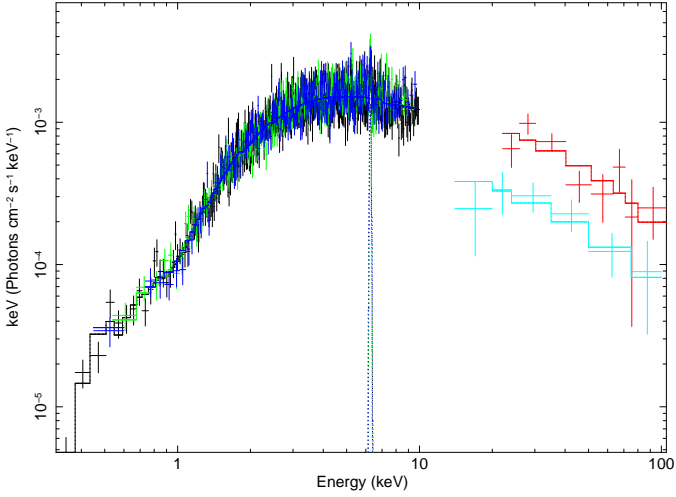}
\caption{0.3--110 keV spectrum of UGC~3142 obtained using non simultaneous EPIC (in the 0.3--10 keV range), BAT (lower flux) and ISGRI (higher flux) data. The fit is a power-law absorbed by cold Galactic matter and by two clumpy neutral absorbers, plus a Gaussian iron K$\alpha$ line (last row of Table \ref{UGC_res_tab}).}
\label{fig:UGC_fin}
\end{figure}%
\subsection{ESO~383-18}
\begin{figure}[t!]
\centering
\includegraphics[width=9cm]{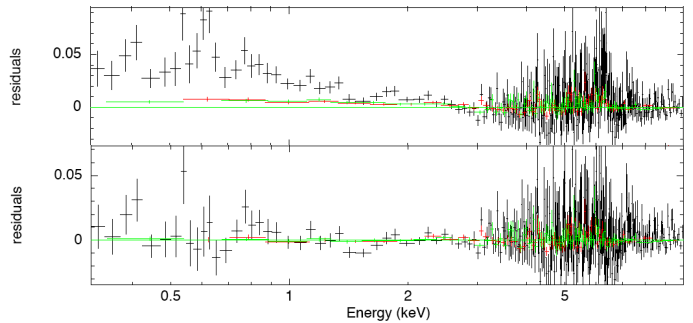}
\caption{ESO~383-18 EPIC/PN (in black) and MOS residuals obtained with the worst (upper panel) and best (lower panel) model, respectively the first and third row of Table \ref{ESO383_res}.}
\label{fig:eso383_res}
\end{figure}

Applying a simple power law absorbed by cold matter to the EPIC spectrum (0.3--10 keV) of ESO~383-18 yields a poor fit ($\chi^{2}$= 1087 for 432 dof), with a low photon index $\Gamma =1.2\pm0.1$. As in the case of UGC~3142, the spectrum of ESO~383-18 presents an excess in the soft X-rays and one around 6.4 keV (Fig.~\ref{fig:eso383_res}), when a simple power law model is applied.

Adding to the power-law continuum a thermal or a non-thermal component, does not represent the emission at low energies ($\chi^{2}$= 930 for 430 dof and $\chi^{2}$= 1051 for 429 dof, for the {\it diskbb} and {\it comptt} model, respectively). 
Applying a power law model absorbed by Galactic cold matter plus a layer of partially absorbing neutral material improves the results at the 99.9\% confidence level ($\chi^2 = 533$ for 431 dof), and steepens the power law ($\Gamma=1.47^{+0.03}_{-0.05}$). Adding another layer of partially absorbing neutral material improves again the result at the 99.9\% confidence level ($\chi^2 = 504$ for 429 dof), and gives a photon index of $\Gamma=1.8^{+0.1}_{-0.1}$. One of the absorbers covers almost completely the X-ray source ($f=0.97^{+0.02}_{-0.04}$) and has a column density of $N_{\rm H}=10_{-2}^{+1} \times 10^{22}\rm \,cm^{-2}$, while the other one has $f=0.86^{+0.05}_{-0.05}$ and a higher column density ($N_{\rm H}=20_{-3}^{+2} \times 10^{22}\rm \,cm^{-2}$). To this model we have added a Gaussian line, which improves the fit again at the 99.9\% confidence level ($\chi^2 = 485$ for 426 dof). This line is consistent with being the iron K$\alpha$ line ($E=6.3^{+0.1}_{-0.1}$ keV) and has a width of $\sigma=23^{+100}_{-32}$ eV, and an equivalent width of EW=$76_{-66}^{+93}$ eV. As for UGC~3142, adding another Gaussian component at 6.7 keV to the neutral iron line does not improve significantly the fit ($\Delta \chi ^2 \simeq 0$). Although the two absorbers have values of column density and covering fraction within the $3\sigma$ range of each other, the second absorber has been proven to be statistically required.

This model still displays an excess between 0.6--0.7 keV in both PN and MOS data. Such an excess could be due to the OVIII Ly$\alpha$ emission line (at 0.65 keV in the rest frame of the source). Implementing this feature in the model as a Gaussian line ($E=0.62_{-0.02}^{+0.02}$ keV) improves the result at the 99\% confidence level ($\chi^2 = 461$ for 423 dof), but the data are not sufficient to constrain significantly the equivalent width of the line: EW=$144_{-137}^{+200}$ eV.

Due to its very low signal to noise ratio the RGS spectrum of ESO~383-18 has not been analyzed.\newline

ESO~383-18 was not detected within the 22-months BAT survey, and the {\it INTEGRAL} IBIS/ISGRI spectrum of ESO~383-18 in the 20-100 keV emission is well represented by a simple power law, with a photon index of $\Gamma=1.9^{+0.4}_{-0.3}$, consistent with the value obtained from the EPIC spectra. Combining the two spectra together and using the best EPIC model we obtain parameters consistent with those obtained previously (Table \ref{ESO383_res}), due to the much richer statistics of EPIC. Due to the low quality of the IBIS/ISGRI data, it is not possible to constrain the presence of a cut-off, or of a reflection component (R$\simeq 0$) in the spectrum using {\it pexrav}. The fit included an inter-calibration constant ($C_{I}=0.7$) in order to account for variability, as ISGRI and EPIC observations were not simultaneous, and the inclination angle was set to $i=60^{\circ}$ (Kinney et al. 2000) as this source is a type 2 AGN.

We can conclude that the best model for the spectrum of ESO~383-18 is a power law absorbed by cold Galactic matter and by a double layer of partially covering material, plus two Gaussian lines.
\begin{figure}[t!]
\centering
\includegraphics[height=9cm,angle=270]{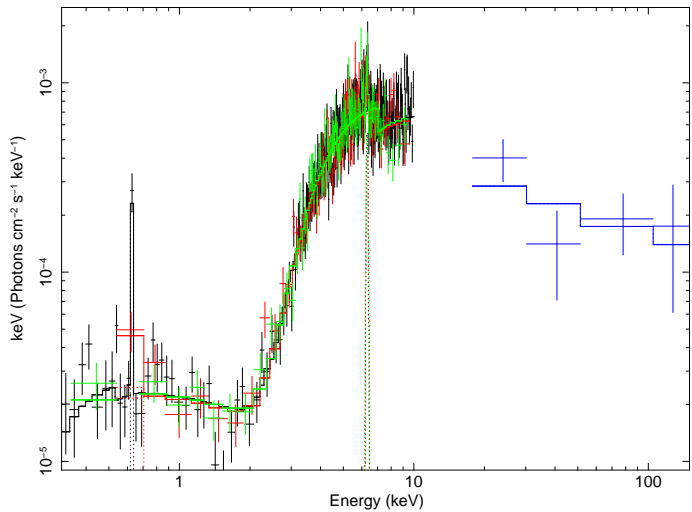}
\caption{0.3--110 keV spectrum of ESO~383-18 obtained using non simultaneous EPIC (in the 0.3--10 keV range) and ISGRI data. The fit is a power law absorbed by cold Galactic matter and by two clumpy neutral absorbers plus two Gaussian emission lines (last row of Table \ref{ESO383_res}).}
\label{fig:eso383_fin}
\end{figure}%
\begin{table*}
\caption[]{Results obtained from the X-ray spectral analysis of ESO~383-18. The data have been modeled using an absorbed power law model ($a$), a power law absorbed by Galactic cold matter plus a layer of partially absorbing neutral material ($b$), and a power law plus 2 Gaussian lines (representing the iron K$\alpha$ and the OVIII Ly$\alpha$ emission lines)s absorbed by Galactic cold matter and by two layers of partially absorbing neutral material ($c$). EPIC includes PN, MOS1 and MOS2 data.}
\label{ESO383_res}	
\begin{tabular}{lllllllll}
\hline
Instruments & Energy&  $\chi ^2$ (dof)& $\Gamma$& $N_{\rm H}$ &$N_{\rm H}^1$&  f$^1$ & $N_{\rm H}^2$ &  f$^2$  \\
& {\scriptsize [keV] }& &  &{\scriptsize [$10^{20}\rm \,cm^{-2}$] }& {\scriptsize [$10^{22}\rm \,cm^{-2}$] }& &{\scriptsize [$10^{22}\rm \,cm^{-2}$] } &  \\
\hline
EPIC$^a$& 0.3--10& 1087 (432)& $1.2\pm0.1$ & $15\pm 1$ &  -- & -- &  -- &  -- \\
EPIC$^b$& 0.3--10&533 (431)& $1.47^{+0.03}_{-0.05}$ & $0.042$ &  $19.5^{+0.7}_{-0.7}$ & $0.989^{+0.002}_{-0.003}$ & -- &  --  \\
EPIC$^c$& 0.3--10& 461 (423)& $ 1.8^{+0.1}_{-0.1}$ & $ 0.042 $ &  $10^{+1}_{-2}$ & $0.97^{+0.02}_{-0.04}$ & $ 20^{+2}_{-3}$ & $ 0.86^{+0.05}_{-0.05}$ \\
%TO FILL!!
(EPIC + ISGRI)$^c$& 0.3--110&  472 (426)& $ 1.7^{+0.2}_{-0.2}$ & $ 0.042 $ &  $11^{+17}_{-5}$ & $0.95^{+0.03}_{-0.15}$ & $ 15^{+4}_{-7}$ & $ 0.85^{+0.06}_{-0.07}$ \\
\hline
\end{tabular}
\end{table*}
%
%%%%%%%%%%%%%%%%%%%%%%%%%%%%%%%%%%%%%%%%%%%%%%%%%%%%%%%%%%%%%%%%%%%%%%%%%%%%%%%%%%%%%%

\subsection{ESO~140-43}\label{ESO140_analysis}
Two {\it XMM-Newton} observations of ESO~140-43 were performed, the first in September 2005 and the second in March 2006.

\begin{figure}[t!]
\centering
\includegraphics[width=9cm]{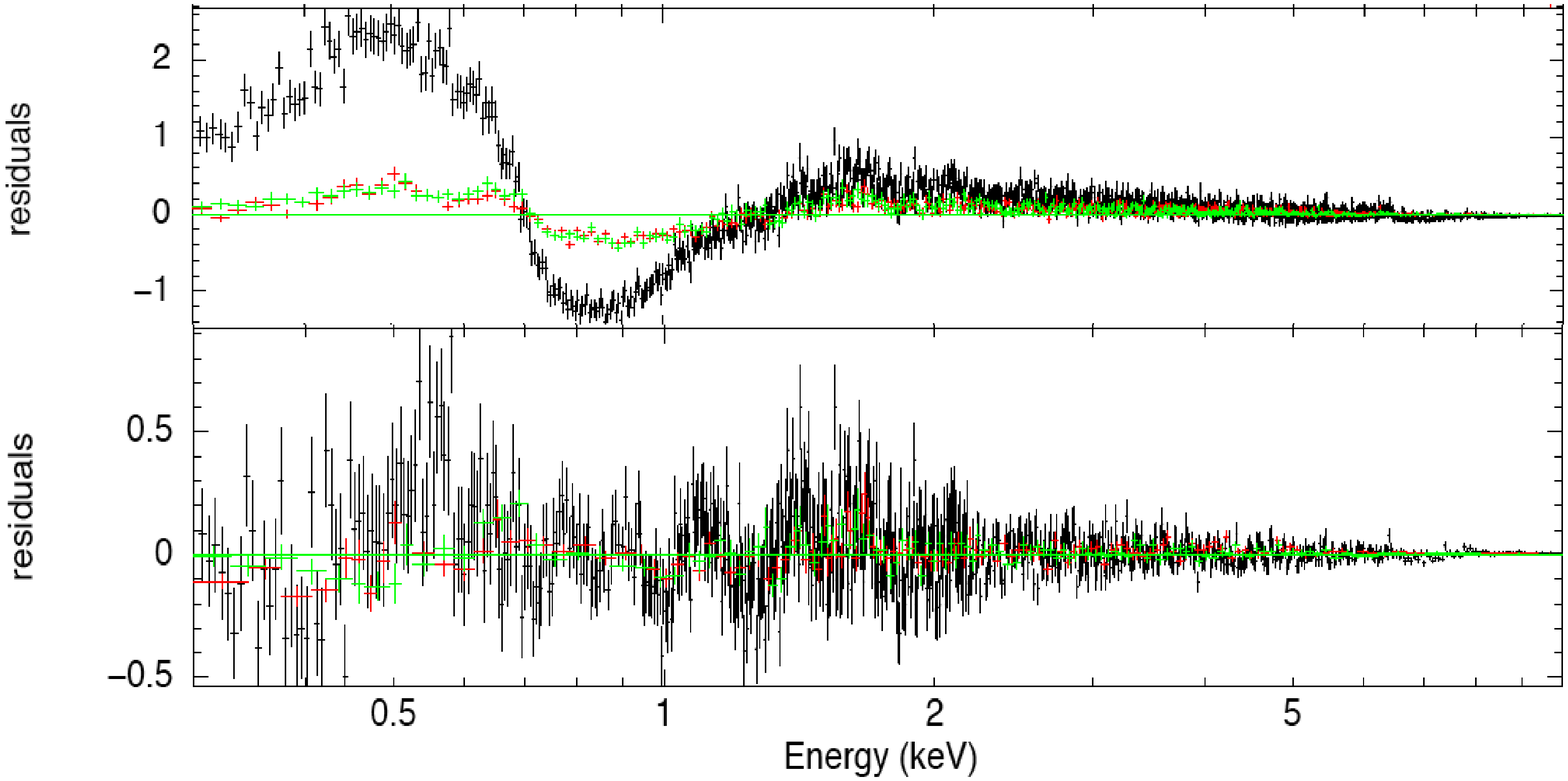}
\caption{EPIC/PN (in black) and MOS residuals of the 2005 {\it XMM-Newton} observation obtained with the worst (simple power law, top panel) and best fit (power law absorbed by cold Galactic matter and by three clumpy ionized absorbers, plus a iron K$\alpha$ line, bottom panel).}
\label{fig:401_res}
\end{figure}

\begin{figure}[t!]
\centering
\includegraphics[width=9cm]{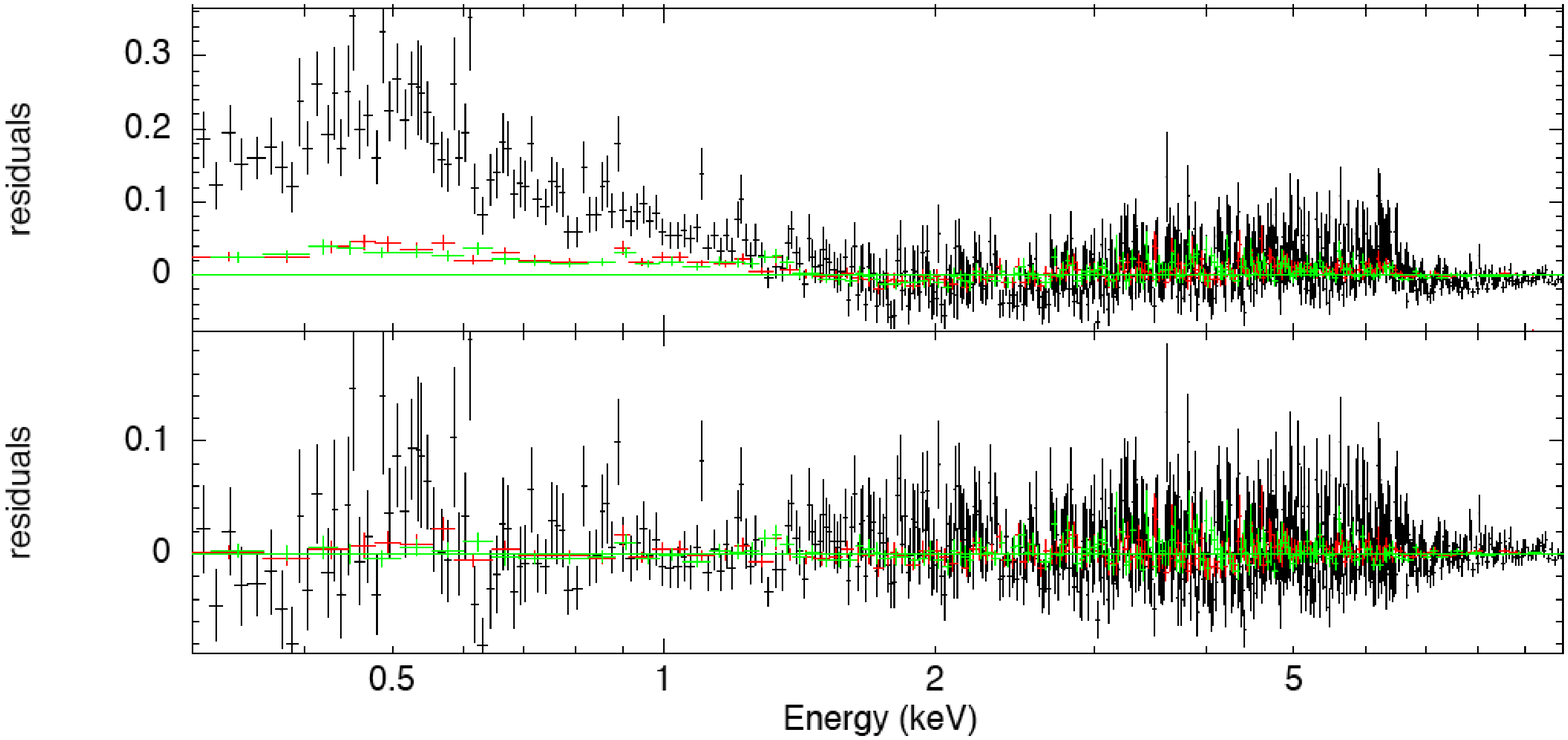}
\caption{EPIC/PN (in black) and MOS residuals of the 2006 {\it XMM-Newton} observation obtained with the worst (simple power law, top panel) and best fit (power law absorbed by cold Galactic matter and by three clumpy ionized absorbers, plus a Gaussian K$\alpha$ iron line, bottom panel).}
\label{fig:901_res}
\end{figure}

As in the case of ESO~383-18 and UGC~3142, in the 0.3--10 keV energy range a simple power law absorbed by cold neutral matter does not provide, for both observations, a good representation of the EPIC/PN and MOS data ($\chi ^2=18,189$ for 2153 dof and $\chi ^2=3,234$ for 947 dof, for the 2005 and 2006 observation, respectively). It also results in a flat power law with a strong soft excess (Fig. \ref{fig:401_res} and \ref{fig:901_res}).
Again we used a more complex cold absorber, but in this case the data are not consistent with neutral absorption. Adopting a clumpy absorber model results in a chi-squared of $\chi ^2=10861$ for 2151 dof for the EPIC spectrum of the 2005-observation. Besides Jim\'{e}nez-Bail\'{o}n et al. (2008) found evidence of absorption features produced by highly ionized material in the RGS spectrum of ESO~140-43, due to 3 warm absorbers in the line of sight.

We analyzed the characteristics of the ionized matter using a model calculated using the XSTAR code (e.g., Kallman et al. 1996, Kallman \& Bautista 2001), which calculates the physical conditions and absorption-emission spectra of photo-ionized gases, considering variable abundances. We used {\it zxipcf}, a model of partial covering absorption by partially ionized material, which uses a grid of XSTAR photo-ionized absorption models. In this model the state of the warm absorber is characterized by the column density of the absorber $N_{\rm H}$, its covering fraction $f$, and the logarithm of the ionization parameter $\xi$, defined as 
\begin{center}
\begin{equation}
\label{eq_absori}
\xi=\frac{L_{ion}}{nr^2},
\end{equation}
\end{center}
where $L_{ion}$ is the isotropic luminosity of the ionizing source in the 5 eV--300 keV energy range, $r$ is the distance of the warm plasma to the source of radiation, and $n$ is the proton density. This model is the ionized equivalent of the {\it pcfabs} model used for UGC~3142 and ESO~383-18.

In the case of the 2005 EPIC observation of ESO~140-43 adding a clumpy warm absorber improves the model ($\chi ^2$= 3409 for 2151 dof) with respect to a simple power law absorbed by cold matter. A second clumpy warm absorber improves again the result at a 99.9\% confidence level ($\chi ^2$=2800 for 2148 dof). The best model is obtained adding a third clumpy warm absorber, which improves the fit again at the 99.9 \% confidence level ($\chi^2$=2575 for 2145 dof). The best model has a photon index of $\Gamma =1.89^{+0.01}_{-0.01}$ and three warm absorbers with quite different characteristics.
The low ionized absorber (hereafter LIA) ($\log\xi =0.28^{+0.05}_{-0.03} \rm \,erg \,cm \,s^{-1}$) with a hydrogen column density of $N_{\rm H}^L=3.7_{-0.7}^{+0.3} \times 10^{21}\rm \,cm^{-2}$ covers a fraction of $ f=0.70^{+0.04}_{-0.04}$ of the X-ray source. 
The intermediate ionized absorber (IIA) ($\log\xi =1.94^{+0.02}_{-0.04} \rm \,erg \,cm\,s^{-1}$) has a higher hydrogen column density of $N_{\rm H}^I=2.99_{-0.01}^{+0.07} \times 10^{22}\rm \,cm^{-2}$ and a covering factor of $ f=0.98^{+0.01}_{-0.05}$. The third absorber has a value of the ionization parameter higher than the other two ($\log\xi =2.73^{+0.06}_{-0.04} \rm \,erg \,cm \,s^{-1}$), and will therefore be called high ionization absorber (HIA). Its hydrogen column density and covering factor are $N_{\rm H}^H=1.1_{-0.2}^{+0.2} \times 10^{22}\rm \,cm^{-2}$ and $ f=0.99^{+0.01}_{-0.01}$, consistent with full coverage. The residuals of this fit (Fig. \ref{fig:401_res}) show that there are still some components in the spectrum which are yet not accounted for, although the model provides an acceptable fit ($\chi^2_\nu = 1.17$). Higher quality data would be needed to investigate whether another absorption component might explain the remaining residuals.

\begin{figure}[t!]
\centering
\includegraphics[angle=270,width=9cm]{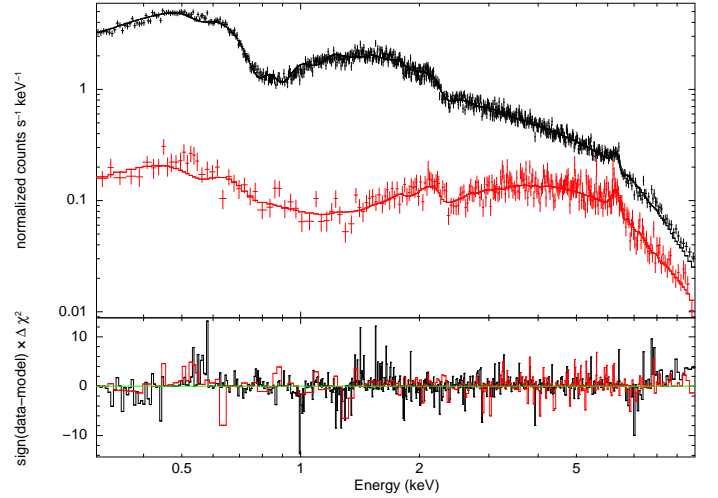}
\caption{EPIC/PN spectra of the 2005 (in black) and 2006 observation of ESO~140-43. The continuous lines represent the best models to the respective data-sets, as reported in Table \ref{ESO140_res_2}.}
\label{fig:401vs901}
\end{figure}

The EPIC/PN and MOS X-ray spectra of the 2006 {\it XMM-Newton} observation of ESO~140-43 were analyzed using the same models reported above. As in the case of  the 2005 observation the best fit ($\chi ^2=1063$ for 940 dof) is obtained using a power law absorbed by three layers of partially covering ionized matter. The value of the photon index obtained ($\Gamma =1.9^{+0.2}_{-0.2}$) is consistent with that of the 2005 observation, while the parameters of the absorbers show several important differences. The LIA has an even lower ionization parameter ($\log\xi ^L =-1.33^{+1.0}_{-0.2} \rm \,erg \,cm \,s^{-1}$) than that of the 2005 observation, it has a higher hydrogen column density ($N_{\rm H}^L=1.6_{-0.8}^{+0.5} \times 10^{22}\rm \,cm^{-2}$) and covers a bigger fraction of the X-ray source ($ f=0.88^{+0.03}_{-0.09}$). 
The IIA presents a value of the ionization parameter ($\log\xi ^I=1.89^{+0.03}_{-0.4} \rm \,erg \,cm \,s^{-1}$) and of the covering factor ($ f=0.86^{+0.09}_{-0.04}$) consistent with the 2005 observation, while its column density ($N_{\rm H}^I=11_{-6}^{+3} \times 10^{22}\rm \,cm^{-2}$) is remarkably higher. 
The HIA has a higher value of the ionization parameter ($\log\xi ^H=3.6^{+0.8}_{-0.2} \rm \,erg \,cm \,s^{-1}$) and of the column density ($N_{\rm H}^H=8_{-6}^{+12} \times 10^{22}\rm \,cm^{-2}$) with respect to the 2005 observation, while its covering factor ($ f=0.86^{+0.09}_{-0.04}$) is consistent.\newline
The spectra of the two observations are reported together in Figure \ref{fig:401vs901}.

A layer of neutral matter representing the Galactic absorption ($N_{\rm H}^{\rm Gal}= 7.34 \times 10^{20}\rm \,cm^{-2}$) has been added to all the models reported above. To discard the presence of neutral absorbers we added a layer of cold material to the models and found that its hydrogen column density is consistent with 0 for both observations, thus it can be ignored.\newline

The first evidence for the presence of a K$\alpha$ iron line in the X-ray spectrum of ESO~140-43 was found by the analysis of EXOSAT data. Ghosh \& Soundararajaperumal (1992) detected an excess around 6 keV, which they interpreted as an iron line of equivalent width EW$=354 \pm 144$ eV.
We added a Gaussian line to the best fit model, and found that the presence of this component improves the fit at the 99.9\% confidence level for both observations ($\chi ^2=2511$ for 2142 dof and $\chi ^2=2012$ for 937 dof, for the 2005 and 2006 observation, respectively). From the spectrum of the 2005-observation we obtained a line located at $6.40^{+0.03}_{-0.03}$ keV, with a width of $\sigma=83^{+34}_{-32}$ eV, an equivalent width of EW=$118^{+51}_{-53}$ eV and a flux of $F_{\rm \,K\alpha}=4.4_{-0.9}^{+0.9}\times 10^{-15}\rm \,erg \,cm^{-2} \,s^{-1}$. In the 2006-observation spectrum the K$\alpha$ iron line is at $6.39^{+0.05}_{-0.05}$ keV, has a width of $\sigma=112^{+57}_{-37}$ eV, an equivalent width of EW=$161^{+10}_{-71}$ eV) and a flux of $F_{\rm \,K\alpha}=2.4_{-0.6}^{+0.7}\times 10^{-15}\rm \,erg \,cm^{-2} \,s^{-1}$. This line for both observations is consistent with being the K$\alpha$ iron line. For both observations adding another Gaussian line at 6.7 kev does not improve significantly the fit ($\Delta \chi ^2 \simeq 0$), which allows us to exclude the presence of an additive broad component to the neutral iron line.
The results obtained using this model are reported in Table \ref{ESO140_res_2}.

\begin{figure}[t!]
\centering
\includegraphics[height=9cm,angle=270]{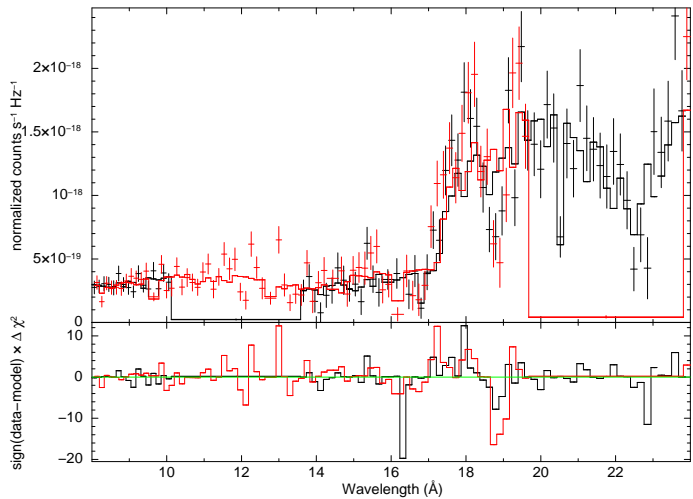}
\caption{Extract of the RGS spectrum of the 2005 observation of ESO~140-43, the solid line represents the best EPIC model (Table \ref{ESO140_res_2}).}
\label{fig:rgs401}
\end{figure}

While the RGS spectrum of the 2005 observation (Fig. \ref{fig:rgs401}) shows a number of interesting features in the 0.3--2 keV energy range, the one obtained during the 2006 observation has a very low signal to noise ratio. The most evident absorption features detected by RGS during the 2005-observation are (the energy are given in the rest frame of the AGN):  the N~VII Ly$\alpha$ (0.5 keV; 24.8 $\AA$), the O~VIII Ly$\alpha$ (0.65 keV; 19 $\AA$) and possibly the Ne~X Ly$\alpha$ (1.02 keV; 12.1 $\AA$) line. In Fig. \ref{fig:rgs401} we plotted the RGS spectrum using the best EPIC model, which fits the continuum well, but not the OVIII Ly$\alpha$ line  and the continuum next to it, similarly to what happens in the best fit of Jim\'{e}nez-Bail\'{o}n et al. (2008). A more detailed analysis with different parameters and possibly other model components is required, but is out of scope for this paper.\newline

We analyzed also the less significant data obtained during the two {\it Swift}/XRT observations. The two observations were performed a few days apart, but due to the fact that the flux varied, we analyzed them individually. Fitting the spectra with a simple power law we obtained $\chi ^2$=18.3 for 15 dof and $\chi ^2$=23 for 14 dof, for the first and second observation, respectively. Due to the low quality of the data it is impossible to resolve the 3 absorbers, thus we applied a simpler model than that used for EPIC data: a power law absorbed by only one layer of clumpy ionized material. This model fits well both observations and the results are reported in Table \ref{ESO140_res_2}.
 
Analyzing IBIS/ISGRI and BAT spectra together with the most significant EPIC spectra, we checked for the presence of a cut-off and/or reflection from neutral material. No evidence of an exponential cut-off was found in the data, the energy of the cut-off reaches, in fact, the upper limit allowed by XSPEC, and it is not constrained ($E_C =500\pm 1000$ keV). Using the $pexrav$ reflection model (fixing the inclination angle to $i=30^{\circ}$ and the cut-off at $E_C = 10,000 \rm \, keV$), we found that the fit improves at the 98\% confidence level ($\chi^2$ = 2540 for 2150 dof and $\chi^2$=2546 for 2151 dof, for the new and the old model, respectively). The value of the reflection obtained is $R=0.6^{+0.1}_{-0.2}$.

We can conclude that the best model for the 0.3--110 keV spectrum of ESO~140-43 is that of a power law reflected by neutral material and absorbed by 3 layers of partially covering ionized material, plus a iron K$\alpha$ line.

\begin{figure}[t!]
\centering
\includegraphics[height=9cm,angle=270]{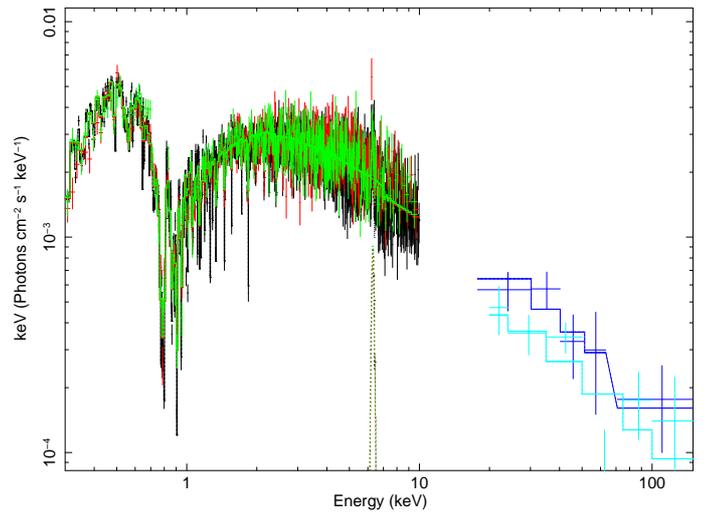}
\caption{0.3--110 keV spectrum of ESO~140-43 obtained using the 2005 {\it XMM-Newton} EPIC (in the 0.3--10 keV range), and non simultaneous BAT (lower flux) and IBIS/ISGRI (higher flux) data. The model used is a power law absorbed by cold Galactic matter and by three clumpy ionized absorbers, plus a iron K$\alpha$ line and a reflection component.}
\label{fig:eso140_401}
\end{figure}%
\begin{figure}[t!]
\centering
\includegraphics[height=9cm,angle=270]{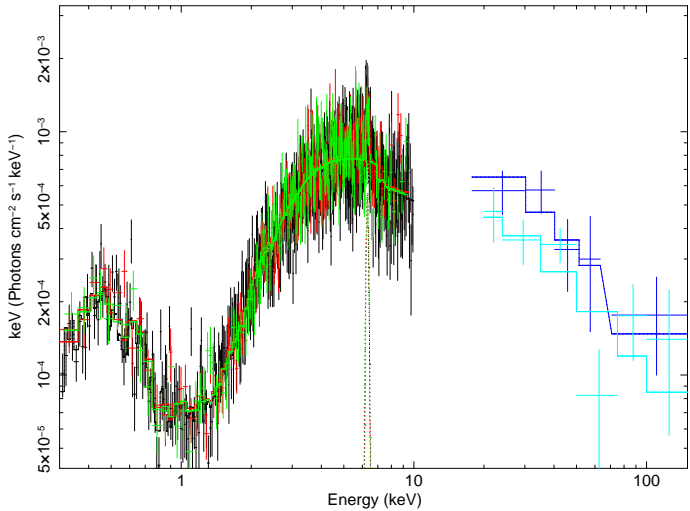}
\caption{0.3--110 keV spectrum of ESO~140-43 obtained using the 2006 {\it XMM-Newton}/EPIC (in the 0.3--10 keV range), and non simultaneous BAT (lower flux) and IBIS/ISGRI (higher flux) data. The model is a power law absorbed by cold Galactic matter and by three clumpy ionized absorbers, plus a iron K$\alpha$ line and a reflection component.}
\label{fig:eso140_901}
\end{figure}%
\begin{table*}
\caption[]{Results obtained from the X-ray spectral analysis of the two {\it XMM-Newton} EPIC and the two {\it Swift}/XRT observations of ESO~140-43. The model used is a power-law plus a iron K$\alpha$ line absorbed by 3 layers of partially covering ionized matter for EPIC, and a power-law absorbed by one layer of partially covering ionized material for XRT. An additional layer of cold matter representing neutral Galactic absorption has been used, fixing the column density at $N_{\rm H}^{\rm Gal}= 7.3 \times 10^{20}\rm \,cm^{-2}$.}
\label{ESO140_res_2}	
\begin{tabular}{lllllllllllll}
\hline
  $\chi ^2$ (dof)& $\Gamma$&$N_{\rm H}^1$& $\log\xi^1$ &  f $^1$&$N_{\rm H}^2$ & $\log\xi^2$  & f $^2$  & $N_{\rm H}^3$ &  $\log\xi^3$  & f $^3$ \\
  &  & {\scriptsize [$10^{22}\rm \,cm^{-2}$] }& {\scriptsize [$\rm \,erg \,cm \,s^{-1}$]}&  &{\scriptsize [$10^{22}\rm \,cm^{-2}$]} & {\scriptsize [$\rm \,erg \,cm \,s^{-1}$] }&  &{\scriptsize [$10^{22}\rm \,cm^{-2}$]} & {\scriptsize [$\rm \,erg \,cm \,s^{-1}$]}  &  \\
  \hline
\multicolumn{11}{l}{\textit{XMM-Newton} EPIC/PN and MOS}\\
 \hline
  & & \multicolumn{3}{|c|}{LIA}& \multicolumn{3}{|c|}{IIA} & \multicolumn{3}{|c}{HIA}\\
  \hline
\multicolumn{11}{l}{\it 2005 Observation}\\
2511(2142) & $1.89^{+0.01}_{-0.01}$&  $0.37^{+0.03}_{-0.07}$ & $0.28^{+0.05}_{-0.03}$ &  $0.70^{+0.04}_{-0.04}$ & $2.99^{+0.07}_{-0.01}$ &$1.94^{+0.02}_{-0.04}$&   $0.98^{+0.01}_{-0.05}$  &  $1.1^{+0.2}_{-0.2}$ & $2.73^{+0.06}_{-0.04}$  &   $0.99^{+0.01}_{-0.01}$ \\
%\hline
\multicolumn{11}{l}{\it 2006 Observation }\\
1012(937) & $1.9^{+0.2}_{-0.2}$& $1.6^{+0.5}_{-0.8}$ & $-1.33^{+1.0}_{-0.2}$ &   $0.88^{+0.03}_{-0.09}$ & $11^{+3}_{-6}$ & $1.89^{+0.03}_{-0.4}$ &  $0.86^{+0.09}_{-0.04}$ & $8^{+12}_{-6}$  &   $3.6^{+0.8}_{-0.2}$ &  $0.99^{+0.01}_{-0.01}$   \\
\hline
\multicolumn{11}{l}{}\\
\multicolumn{11}{l}{\textit{Swift}/XRT}\\
\hline
\multicolumn{11}{l}{\it First observation}\\
13.1(13) & $3.8^{+1.0}_{-2.2}$&  $7^{+13}_{-6}$ & $-0.5^{+2.3}_{-0.7}$ & $0.99^{+0.01}_{-0.01}$  & & & & & &  \\
%\hline
\multicolumn{11}{l}{\it Second observation}\\
12.9(12) & $2.9^{+1.4}_{-1.2}$& $5^{+2}_{-1}$ & $-0.5^{+2.3}_{-0.1}$  &  $0.996^{+0.004}_{-0.03}$ & & & & & &\\
\hline
\end{tabular}
\end{table*}
\begin{table*}
\caption[]{Optical and ultraviolet apparent magnitudes measured by the two different {\it XMM-Newton}/OM and {\it Swift}/XRT observations. }
\label{eso_140_UV}	
\begin{center}
\begin{tabular}{lccccc}
\hline
Instrument & & XMM/OM & XMM/OM & Swift/UVOT & Swift/UVOT \\
 Time & &  08-09-2005 & 07-03-2006 & 20-05-2008 & 25-05-2008\\
 \hline
 Band & Wavelength & Magnitude  & Magnitude & Magnitude & Magnitude\\
 & {\footnotesize $\AA $}& {\footnotesize mag} & {\footnotesize mag} & {\footnotesize mag}& {\footnotesize mag}  \\
\hline
V & 5430 &$ 14.56^{+0.01}_{-0.01}$ & $ 14.57^{+0.01}_{-0.01}$  & -- & -- \\
B & 4500 &$ 15.34 ^{+0.01}_{-0.01}$ & $ 15.37^{+0.01}_{-0.01}$ & -- & -- \\
U & 3440 &$ 14.90 ^{+0.01}_{-0.01}$  & $ 14.94 ^{+0.01}_{-0.01}$ & -- & -- \\
UVW1 & 2910 &$ 15.14^{+0.01}_{-0.01}$ & $ 15.18^{+0.01}_{-0.01}$  & -- & $ 16.0^{+0.1}_{-0.1}$ \\  
UVM2 & 2310  &$ 16.14 ^{+0.04}_{-0.04}$ & $ 16.67^{+0.04}_{-0.04}$  & $ 16.8^{+0.1}_{-0.1}$ & -- \\
$\rm UVW2$ & 2120 &$ 16.57 ^{+0.08}_{-0.08}$ &  -- & -- &--\\
\hline
\end{tabular}
\end{center}
\end{table*}%
\begin{table*}
\caption[]{Comparison between fluxes in different energy bands obtained in this work and found in the literature. The indices $1$ and $2$ refer to the 2005 and 2006 {\it XMM-Newton} observation of ESO~140-43, respectively. The fluxes are model fluxes, not corrected for absorption.}
\label{tab_fluxes}	
\begin{center}
\begin{tabular}{lllll}
\hline
Instrument &  $F_{0.1-2 \rm \,keV}$ & $F_{0.1-2.4 \rm \,keV}$ & $F_{2-10 \rm \,keV}$& $F_{14-150 \rm \,keV}$\\
 \hline
 & {\scriptsize  [$10^{-12}\rm \,erg \,cm^{-2}\,s^{-1}$]} & {\scriptsize [$10^{-12}\rm \,erg \,cm^{-2}\,s^{-1}$]} & {\scriptsize [$10^{-11}\rm \,erg \,cm^{-2}\,s^{-1}$]} & {\scriptsize [$10^{-11}\rm \,erg \,cm^{-2}\,s^{-1}$]}  \\
\hline
 \multicolumn{5}{l}{UGC~3142}\\
{\it EPIC}/PN + MOS & $0.69^{+0.04}_{-0.11}$ & $1.24^{+0.1}_{-0.1}$ & $1.7^{+0.1}_{-0.2}$ & -- \\
{\it Swift}/BAT $^a$&  -- & -- & --& $3.4 ^{+0.6}_{-0.6}$ \\
{\it INTEGRAL} IBIS/ISGRI&  --& --& --& $9.42^{+0.6}_{-0.8}$ \\
\hline
 \multicolumn{5}{l}{ESO~383-18}\\
{\it EPIC}/PN + MOS & $0.06^{+0.18}_{-0.01}$ & $0.08^{+0.18}_{-0.01}$ & $0.6^{+0.1}_{-0.1 }$ & --  \\
{\it Swift}/BAT $^b$& -- & -- & -- & $1.9 ^{+0.6}_{-0.6}$ \\
{\it INTEGRAL} IBIS/ISGRI& -- & -- & -- & $3.8^{+1.8}_{-1.5}$ \\
\hline
 \multicolumn{5}{l}{ESO~140-43}\\
{\it ROSAT} $^c$ &  -- & $21^{+10}_{-10}$ &-- &--  \\
{\it EPIC}/PN + MOS$^1$ & $6.5^{+0.1}_{-0.1}$ & $8.5^{+1.1}_{-0.6}$& $2.46^{+0.05}_{-0.04}$ & --  \\
{\it EPIC}/PN + MOS$^2$ &  $0.4^{+0.1}_{-0.1}$ &$0.5^{+0.2}_{-0.2}$ & $0.8^{+0.2}_{-0.2}$ & --  \\
{\it EXOSAT} $^d$ &  $8^{+3}_{-3}$ &--& $2.5^{+0.2}_{-0.2}$ & -- \\
{\it Swift}/BAT $^a$&  -- & -- & -- & $4.6^{+0.7}_{-0.7}$  \\
{\it INTEGRAL} IBIS/ISGRI& -- & -- &-- & $5.3^{+0.4}_{-0.5}$ \\
\hline
\multicolumn{5}{l}{$^a$ Tueller et al. 2010, $^b$ Cusumano et al. 2009, $^c$ Boller et al. 1992, $^d$ Ghosh \& Soundararajaperumal 1992.}\\
\end{tabular}
\end{center}
\end{table*}%
\section{Variability analysis}\label{X_UV_variability}
\subsection{X-ray variability}\label{soft_var}
\begin{figure*}[t!]
\centering
 %% 1st image
\begin{minipage}[!b]{.48\textwidth}
\centering
\includegraphics[height=2.8cm]{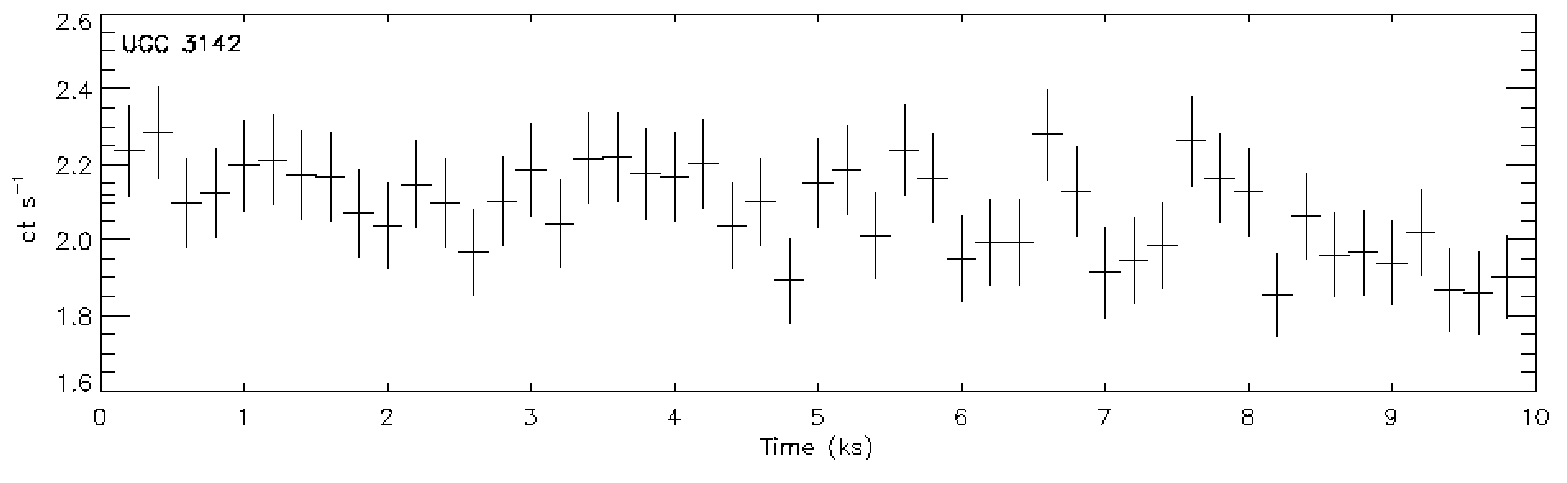}
\end{minipage}
\hspace{0.05cm}
 %% 2nd image
\begin{minipage}[!b]{.48\textwidth}
\centering
 \includegraphics[height=2.8cm]{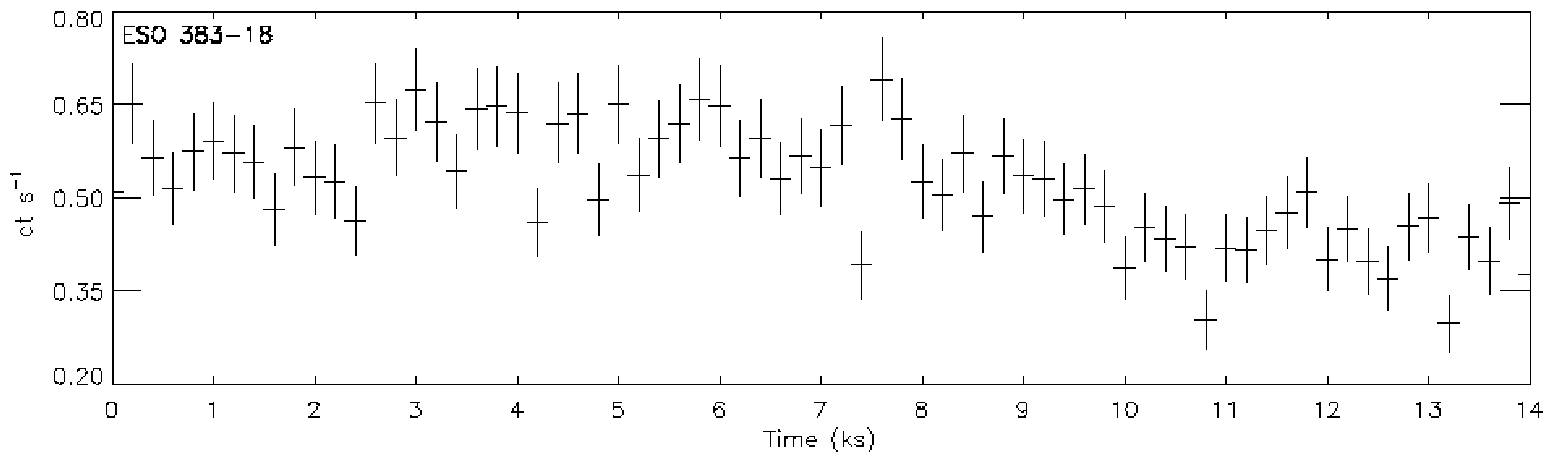}
 \end{minipage}
% %% caption
 %% 1st image
\begin{minipage}[!b]{.48\textwidth}
\centering
\includegraphics[height=2.8cm]{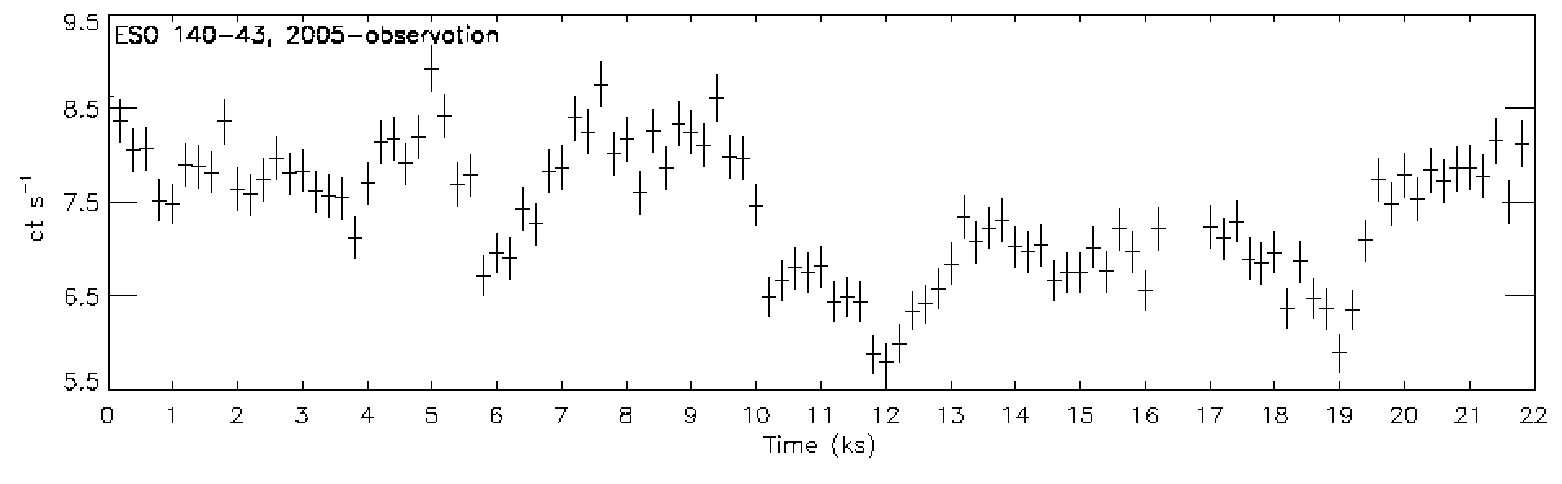}
\end{minipage}
\hspace{0.05cm}
 %% 2nd image
\begin{minipage}[!b]{.48\textwidth}
\centering
 \includegraphics[height=2.8cm]{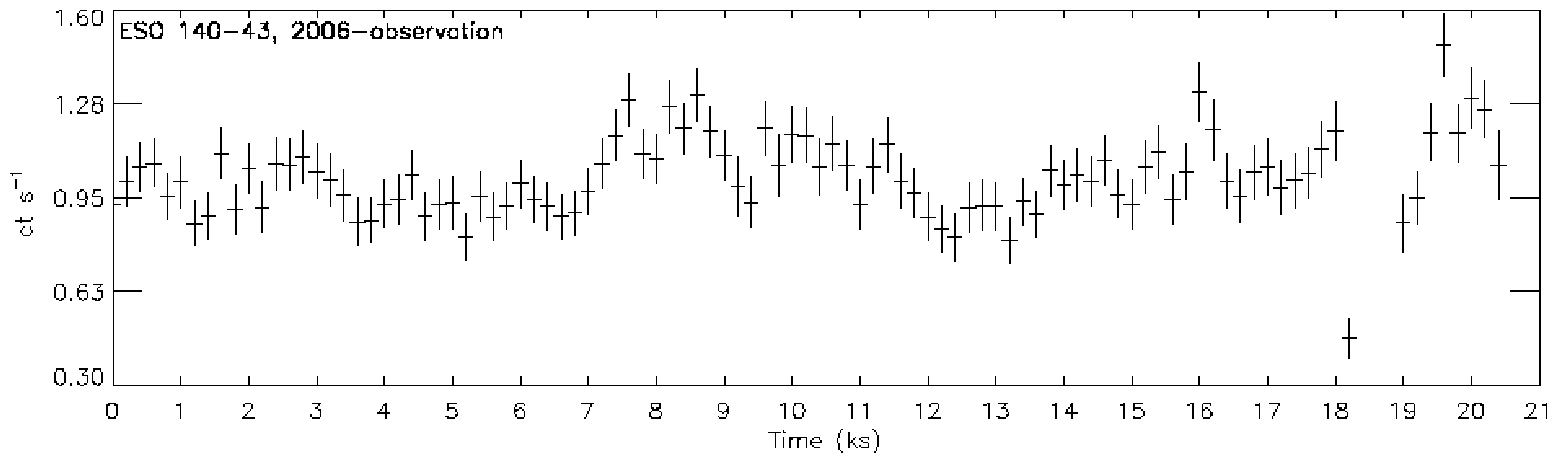}
 \end{minipage}
% %% caption
 \begin{minipage}[t]{1\textwidth}
  \caption{EPIC PN light curves binned to 200 seconds for UGC~3142 (top left panel), ESO~383-18 (top right panel) and the 2005 (bottom left panel)  and the 2006 observation of  ESO~140-43 (bottom right panel).}
\label{fig:lc}
 \end{minipage}
\end{figure*}

We analyzed the 0.3--10 keV EPIC/PN light curves (with a binning of 200s) of the 3 Seyfert galaxies (Fig. \ref{fig:lc}) and investigated whether variations in the absorbers can account for variability. We tested the light curves for a constant flux model, and rejected the null hypothesis (i.e. flux not variable) for values of the probability $\rm \,p$ lower than $5\%$.

The light curve of UGC~3142 is consistent with a constant flux ($\chi ^2= 54$ for 49 dof, which leads to a probability of $p\sim 30\%$), with a mean count-rate of $ \bar f= 2.08\pm 0.02 \rm \,ct \,s^{-1}$.
The light curve of ESO~383-18 shows significant variability during the observation, with $\chi ^2= 187.5$ for 70 dof ($p\sim 0\%$) for the assumption of constant flux.
In order to better estimate the variability of these two light-curves we calculated their fractional rms variability amplitude $F_{var}$ (as defined in Eq. 10 in Vaughan et al. 2003). For UGC~3142 we obtained $F_{var}=1.5\pm 2\%$, compatible with the lack of variability found above, while for ESO~383-18 we obtained $F_{var}=14\pm 2\%$. The errors on $F_{var}$ were calculated using Eq. B2 in Vaughan et al. (2003).

The light curves for the two observations of ESO~140-43 show evidence of variable flux,  with $\chi ^2= 1575$ for 107 dof ($p\sim 0\%$) and $\chi ^2= 257$ for 99 dof ($p\sim 0\%$), for the 2005 and 2006 observations, respectively. The values of the fractional rms variability amplitude are consistent, $F_{var}=8.9\pm 0.3\%$ and $F_{var}=9\pm 1\%$, for the 2005 and 2006 observation, respectively.

\begin{figure}[t]
\centering
\includegraphics[height=9cm,angle=270]{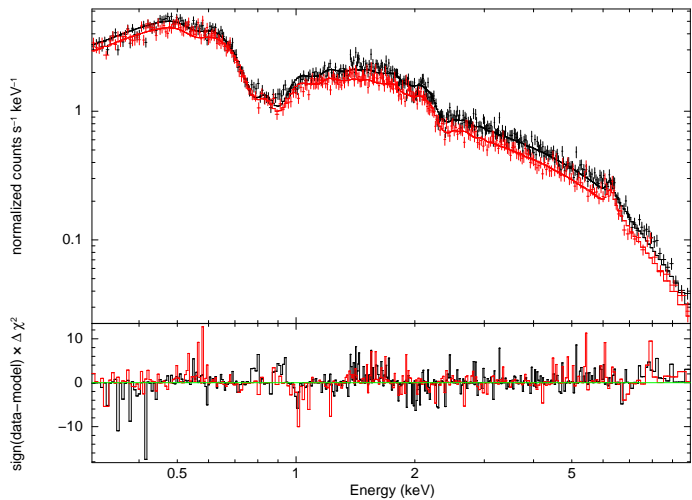}
\caption{ EPIC/PN spectra of different time intervals according to the low flux state and the high flux state (in black) during the 2005-observation of ESO~140-43. The model is the best model obtained in Section~\ref{ESO140_analysis}, with all parameters fixed to the best model of this observation, except the covering factor of the LIA and the normalization.} %, which were left free to vary.}
\label{fig:pndivided}
\end{figure}

The 2005-observation seems to be characterized by two different (at the 1$\sigma$ level) flux states. The light curve shows in fact a constant flux of about 8 ct/s for the first 10 ks, then a dip and a flux oscillating around 6.7 ct/s for about 9 ks, then increasing again up to 10 ct/s. In order to test whether the different flux states can be accounted for by a change in the absorber or in the spectral shape, we divided the light curve into two parts: {\it part 1}, between 0 and 10 ks for the high flux state, and {\it part 2}, between 10 and 19 ks, for the low flux state. These two light-curves have consistent intrinsic variabilities of few percents ($F_{var}^1=5.1\pm 0.4\%$ and $F_{var}^2=5.2\pm 0.5\%$). In order to find a possible explanation of the existence of these two flux states, we analyzed the spectra obtained from the two time intervals (Fig. \ref{fig:pndivided}). We fitted the two spectra first freezing all the parameters except the normalization of the power-law, and then letting also the other parameters free to vary. We explored several possibilities, as variations of the column densities, of the ionization parameters or of the covering fraction. We found that the only parameter which variation could explain the different spectra, together with the normalization, is the covering factor of the LIA. In particular a variation of this parameter (from $ f^{LIA}_{1}=0.77^{+0.01}_{-0.02}$ to $ f^{LIA}_{2}=0.65^{+0.02}_{-0.03}$) is necessary at the 99.9\% confidence level to account for these spectral differences.

In Table \ref{tab_fluxes} we report the fluxes of the 3 sources based both on this work and previous literature. A strong variation in flux in the 14--150 keV band is evident between ISGRI and BAT observations of UGC~3142, while the values of ESO~383-18 and ESO~140-43 are consistent, and there is no evidence of hard X-ray variability. At lower energies flux (by a factor of 3) and spectral variability on a time-scale of months was detected between the two {\it XMM-Newton} observations of ESO~140-43. The fluxes of ESO~140-43 measured by {\it EXOSAT} are consistent with those obtained by the 2005 EPIC observation, while the flux was much lower during the 2006 observation. Strong variability between the {\it ROSAT} observation and the second EPIC observation was also detected.\newline
Thus of the three Seyferts only UGC~3142 shows strong hard X-ray variability, while in soft X-ray ESO~383-18 and ESO~140-43 show both variability on the 200s timescale, and ESO~140-43 shows remarkable flux and spectral variability on a monthly timescale, as evident from {\it XMM-Newton} and historical observations.

%%%%
\subsection{Hardness ratios}\label{HR_var}
In order to probe the possible role of absorption in the 0.3--10 keV flux variability of ESO~383-18 and ESO~140-43, we used two bands, a soft ({\it S}, 0.3--2 keV) band and a hard ({\it H}, 2--10 keV) band, and checked for variations of the hardness ratio with time.
The hardness ratio is defined as:
\begin{equation}
\label{HR}
HR = \frac{H-S}{H+S},
\end{equation}
where {\it S} and {\it H} are in $\rm \,counts \,s^{-1}$. We searched for the presence of correlations using Pearson's product moment correlation coefficient {\it r}, defined as
\begin{equation}
\label{Pearson}
r=\frac{1}{n-1}\sum_{i=1}^n\left (\frac{x_i-\bar{x}}{\sigma_{x}}\right ) \left (\frac{y_i-\bar{y}}{\sigma_{y}}\right ),
\end{equation}
where $x_i$ and $y_i$ are the values of the two data-sets we are comparing (namely the HR and the flux or the time), $n$ the size of the samples, $\bar{x}$ and $\bar{y}$ are the averages of the data-sets, and $\sigma_{x}$ and $\sigma_{y}$ are their standard deviations. In order to calculate the probability the null hypothesis ($r=0$, namely no correlations) is true, we used the Fisher transformation 
\begin{equation}
\label{Fisher}
F(r)=\frac{1}{2}\log\frac{1+r}{1-r}
\end{equation}
from which we calculated $z= \sqrt{n-3}F(r)$, a z-score of $r$, from which we derived the probability $p$. We rejected the null hypothesis in the case $p<5\%$.

The hardness ratio of ESO~383-18 is not constant ($\chi^2 = 34$ for 70 dof, $p=0.03\%$), but does not show any significant correlation with time ($r=-0.3$, $p=99\%$) or flux ($r=0.05$, $p=65\%$). This lack of correlations might imply that the observed spectral variability on this time-scale is not due to absorbing material.

During the 2005 observation of ESO~140-43 the hardness ratio varied ($\chi^{2}=188$ for 107 dof, $p<0.01\%$), while it is consistent with no variability during the 2006 observation ($\chi^{2}=24.3$ for 99 dof, $p\simeq 100\%$). The 2005 observation does not show any correlation of the hardness ratio with flux ($r= 0.26$, $p= 61\%$) or time ($r= -0.55$, $p= 99.9\%$), which indicates that changes in the absorbing material are not a likely explanation for the spectral variability on the 200s time-scale. A strong variation of the average value of HR appears between the 2005 ($\overline{\rm \,HR^1} =-0.186 \pm 0.003$) and the 2006 ($\overline{\rm \,HR^2} = +0.53 \pm 0.02$) observations, consistent with the observed spectral variation.

\subsection{Optical and UV variability}\label{uv_var}
We analyzed the optical and UV emission of ESO~140-43 at the time of the X-ray observations, using {\it XMM-Newton}/OM and {\it Swift}/UVOT data. In Table~\ref{eso_140_UV} the values of optical and ultraviolet magnitudes obtained are given. In the V, B, U, and UVW1 band the fluxes obtained by the two {\it XMM-Newton} observations are consistent, while in the UVM2 band the source varied by about $\Delta \rm UVM2  \simeq 0.5$ mag. UVOT measured the flux only in the UVW1 and UVM2 bands, and registered strong flux variations in UVM1 with respect to the OM observations ($\Delta \rm UVW1 \simeq 0.9$ mag), while the UVM2 flux remained constant.
%%%%
\section{Discussion}\label{discussion}
In this work we presented two AGN showing evidence for clumpy neutral absorbers and one showing evidence for clumpy ionized absorbers. We discuss these two cases separately, and then together in light of the soft X-ray excess.

\subsection{UGC~3142 and ESO~383-18}
Recent results (e.g., Risaliti et al. 2009, Bianchi et al. 2009) show that clumpy absorbers can explain variability in AGN. Both, Seyfert 1 (i.e. Guainazzi et al. 1998,  Puccetti et al. 2007, Grupe et al. 2007) and Seyfert 2 (Risaliti et al. 2002) have been proven to show partial or total occultation events. In particular several Seyfert 2 have been observed changing state, switching from Compton-thin to reflection-dominated or vice versa (Matt et al. 2009). In their recent work Bianchi et al. (2009) propose a model consisting of 3 absorbers/emitters, not necessarily coexisting or observable in all the sources. These 3 components are a Compton-thick torus (at about parsec distance), a Compton-thin absorber ($N_{\rm H}\sim 10^{22}\rm \,cm^{-2}$) at much larger scales, likely associated to dust lanes, which completely or partially obscures the broad line region (BLR), and clouds at around the distance of the BLR. Recent evidence of clumpy structures in the torus has been found from {\it Spitzer}/IRS spectra (Mor et al. 2009). A clumpy moving absorber could also explain the optical change in the Seyfert type observed in some AGN (e.g., in Cohen et al. 1986, Aretxaga et al. 1999).

Both UGC~3142 and ESO~383-18 spectra show what at first sight appears to be a soft X-ray excess, though our analysis shows that the best fit to the data is obtained using {a power law absorbed by two clumpy neutral absorbers}. For the Seyfert~1 UGC~3142 we found that the two absorbers have column densities of the order of $10^{22}\rm \,cm^{-2}$, and covering fractions of about 92\% and 60\%.
In the case of the Seyert~2 ESO~383-18, as expected from its optical classification, the column densities are larger, of the order of $10^{23}\rm \,cm^{-2}$, with covering fractions of about 97\% and 86\%.
The intrinsic absortion of the Seyfert~1 galaxy UGC~3142 is above the dividing line of $10^{22} \rm \,cm^{-2}$ which usually points to a type 2 AGN. It has to be taken into account, though, that the transition between absorbed and unabsorbed sources is smooth, and that not all type 1 AGN necessarily exhibit low absorption (Awaki et al. 1991). In addition, Seyfert~2 galaxies like NGC~3147 and NGC~4698 show no intrinsic absorption (Pappa et al. 2001). The {\it INTEGRAL} AGN catalogue (Beckmann et al. 2009) contains 9 type 1 AGN with $N_{\rm H} > 10^{22} \rm \, cm^{-2}$ and 10 Seyfert 2 with $N_{\rm H} < 10^{22}$ out of 154 Seyfert galaxies with known intrinsic absorption.

Following the model of Bianchi et al. (2009), the presence of the partially covering absorbers seen in ESO~383-18 and UGC~3142 implies a certain degree of clumpiness in the BLR or in the torus, or the existence of dust lanes. An additional component might be represented by a clumpy neutral (or weakly ionized) disc wind, crossing the line of sight. These outflows, originating in the accretion flow, are predicted by hydrodynamic simulations of accretions disks (e.g., Schurch et al. 2009).
According to this, in the case of ESO~383-18 we could picture the absorption as due to a clumpy torus, plus a disc wind or Compton thin dust lanes with a smaller column density, while in the case of UGC~3142, we could make the hypothesis that the observed absorbers are part of a disc wind or of the BLR.

Variations in the partial covering absorbers have been shown to cause strong spectral and flux variability in AGN (e.g., 1H~0707-495, Gallo et al. 2004). ESO~383-18 shows variability in both flux (of $\simeq 14 \%$) and hardness ratio, but, as the hardness ratio is not correlated with time or flux, we exclude that the changes on the 200 seconds time-scale might be caused by changes in the absorbers. The observed spectral variability is probably due to changes of the continuum.
In the EPIC spectrum of ESO~383-18 we detected a feature that could be associated with the OVIII Ly$\alpha$ emission line, but due to the low significance of the data this has to be considered as a tentative identification. The OVIII Ly$\alpha$ appears to be a common feature in absorbed AGN, and it was detected in 30 of the 69 nearby AGN studied by Guainazzi \& Bianchi (2007).

For both objects we cannot confirm or rule out the hypothesis that the neutral matter responsible for clumpy absorption is part of a disc wind. In fact, blue-shifted absorption lines, which would indicate out-flowing matter, are not detected. We can nevertheless safely state that the iron K$\alpha$ line in both objects is not produced in out-flowing material, since the line's centroid is consistent with the redshift of the host galaxy. The non-detection of a reflection hump in the two AGN, despite the presence of narrow neutral iron lines, is probably due to the low significance of IBIS/ISGRI and BAT data for ESO~383-18, and might be hidden by the strong variability in hard X-rays for UGC~3142.

It should be taken into account that partial-covering models are indistinguishable from Thomson scattering models (Turner \& Miller 2009), thus an alternative explanation could be that part of the continuum has been scattered energy-independently and that the absorbers are not partially covering the X-ray source.

\subsection{ESO~140-43}
Warm absorbers have been found to be a common characteristic of Seyfert~1 AGN. In a study of 15 Seyfert~1 McKernan et~al. (2007) found that 10 sources of their sample present out-flowing photo-ionized absorbers, with velocities in the range $\rm \,v \sim 0-2000 \rm \,km \,s^{-1}$, column densities of about $N_{\rm \,H}\sim10^{20-23}\rm \,cm^{-2}$ and ionization parameters in a wide range of $\xi\sim 1-10^4 \rm \,erg \,cm \,s^{-1}$. Several of these absorbers show multiple ionization and kinematics components.

The Seyfert~1 galaxy ESO~140-43 shows a very complex X-ray spectrum, which can be modeled using a power law absorbed by 3 layers of partially covering ionized material, plus a Gaussian iron K$\alpha$ line at the same redshift as the host galaxy, and a weak reflection hump. The {\it XMM-Newton}/RGS data of one of the two observations of ESO~140-43 were analyzed by Jim\'{e}nez-Bail\'{o}n et al. (2008), who also detected 3 warm absorbers, and, contrarily to what we obtained here, also found an additive cold absorber. The photon index they obtained ($\Gamma=2.0\pm0.1$) is consistent with what was found here ($\Gamma \simeq 1.9$). The results we obtained are in agreement with the current picture of multi-structure warm absorbers of Seyfert~1 AGN, in which several zones of warm material with different ionization states are observed.  Examples are the Seyfert~1 galaxies NGC~3783 (Netzer et al. 2003, Krongold et al. 2003) and MCG~6-30-15 (McKernan et al. 2007), the last of these showing evidence for a similar complex absorber consisting of three zones of ionized gas. The Seyfert 1 NGC~3516 (Turner et al. 2005) shows evidence for a warm absorber covering about 50\% of the continuum source, with the variation of its covering factor being used to explain the observed spectral variability (Turner et al. 2008). A powerful tool for the diagnostic of the physical conditions of ionized material in AGN is the study of absorption lines below 2 keV (e.g., Porquet \& Dubau 2000). The RGS spectrum of ESO~140-43 shows the presence of several absorption lines due to H- and He-like elements. These lines are a common characteristics of Seyfert 1 galaxies (e.g., Longinotti et al. 2008, Armentrout et al. 2007, Nucita et al. 2010) and are also found in several absorbed AGN (Bianchi \& Guainazzi 2007). The best X-ray model, the presence of absorption lines due to ionized material and the strong variability of ESO~140-43 imply the existence of ionized clumpy absorbers, which cross the line of sight toward the central engine producing the X-ray continuum.

Due to the strong spectral and flux variability on a time-scale of months, the 3 clumpy warm absorbers detected in the spectrum of ESO~140-43 could be part of a disc wind, as seen in several other Seyfert~1 (e.g., Turner et al. 2008). But, as we do not detect any evidence of outflows, we cannot confirm this hypothesis. Possible explanations of the mechanism by which the out-flowing gas is accelerated are thermal winds (Krolik \& Begelman 1986), radiation pressure (Murray et al. 1995), or magnetically driven winds (Bottorff et al. 2000). In the disc wind scenario the observed variability could be due to changes in the physical conditions of the disc, or caused by interaction of a wind with other material. The detection of UV variability, considering that the optical/UV emission is probably disc dominated, supports the assumption that the observed variability is due to changes in the physical conditions of the accretion disc. An alternative scenario might be that the 3 warm absorbers are part of the BLR, and we observe occultations, as seen in the case of NGC 1365 (Risaliti et al. 2007).
Both EPIC observations of ESO~140-43 show a narrow neutral iron line, and no evidence of a broad component. The line flux varied by about 45\% between the two observations, comparable with the variation of the continuum. As all the absorbers we detected are ionized, we can conclude that the neutral iron line and the reflection hump (which is also due to reflection on neutral matter) are produced by material which is not located in the line of sight. Possible locations for the line emitting region might be the torus or the outer part of the accretion disc.

Flux (of $\simeq 9\%$) and spectral variability on the time-scale of $200 \rm \, s$ is observed in the light curves obtained from both EPIC observations, but the lack of significant correlations between the hardness ratio and the flux indicates that the absorbers do not influence strongly the variability on this time-scale, and that variations in the continuum probably play a more important role. On longer time-scales (kiloseconds) the influence of the absorbers starts to be more evident. A variation of the covering factor of the LIA is in fact necessary to model the change between the high and the low flux state of the 2005 observation. This result is in agreement with the strong influence of the absorbers on the observed spectral variability of ESO~140-43.

The location and geometry of the ionized gas are still debated, but a rough estimate of the upper limit on its distance to the ionizing source can be given (Turner et  al. 2008) by
\begin{equation}
\label{eq_dist}
%\begin{center}
r \lesssim \frac{L}{\xi N_{\rm \,H}}.
%\end{center}
\end{equation}
Using this relation, and considering the values obtained by the best fitting models, we estimated upper limits for the distance to the central engine of the 3 ionized absorbers. For the 2005 observation we obtained $r_{HIA}\lesssim  2.9$ pc, $r_{IIA}\lesssim 6.6$ pc and $r_{LIA}\lesssim 2.5$ kpc, while for the 2006 observation $r_{HIA}\lesssim  0.05$ pc, $r_{IIA}\lesssim 2$ pc and $r_{LIA}\lesssim 23$ kpc. The luminosity has been calculated extrapolating the power-law continuum into the 5 eV -- 300 keV band. These values are not constraining, except for the case of the HIA, which appears to be significantly closer to the central engine than the torus. The value of the HIA obtained from the 2006 observation is similar to what has been found for the highest ionization absorber of NGC~3783 (Reeves et al. 2004). All the upper limits found here are well above the value of the inner radius of the BLR of ESO~140-43 found by Padovani \& Rafanelli (1988), $r^{\rm \,BLR}_{in}\simeq 8.9 \times 10^{-3}$ pc. This result does not allow us to exclude the BLR to be, at least partially, responsible for the absorption.

\subsection{Clumpy absorbers and the soft excess}
{The nature of the soft X-ray excess of AGN has remained a mystery for the past 2 decades. Several explanations have been proposed, here we focus in particular on the absorption models (e.g., Gierlinski \& Done 2004, Guainazzi et al. 2005, Schurch \& Done 2006). In the velocity smeared, partially ionized absorption scenario (Gierlinski \& Done 2004, Schurch \& Done 2006) the observed soft excess is related to atomic transitions, in particular to the large increase in opacity in partially ionized material, due to lines and edges of ionized OVII, OVIII and iron at $\sim$0.7 keV. In this model, the partially ionized material is thought to be a disc wind, with a complex velocity structure, which makes it difficult to resolve the atomic features. Schurch \& Done (2008) noted that magnetic driving is likely the only mechanism that could achieve the high velocities required. While we do not detect any evidence of high velocities components in the absorbing medium, we found that the absorbers are clumpy, consistent with the assumptions of Schurch \& Done (2008). 

Warm absorbers are likely to be the main cause for the soft X-ray excess observed in many AGN (Turner et al. 2009), in this context ESO~140-43, with its 3 partially covering ionized absorbers, gives further evidence on how absorption can explain the soft excess in some Seyfert galaxies.
Neutral absorbers can also play a significant role in the observed soft excess of AGN (Matt et al. 2001, Guainazzi et al. 2005). In the clumpy neutral absorber scenario adopted here, the X-ray source is not completely covered by the absorbers, and this allows part of the flux below 2 keV to leak out. Considering an absorber totally obscuring the primary X-ray emission (i.e. the power law continuum), we over-estimate the absorption below 2 keV, which then results in an apparent excess over the model.

Although for the 3 AGN presented here the absorption model fits well the observed soft X-ray excess, from previous studies it is obvious that not one model alone can account for what is observed in the many sources. Guainazzi et al. (2005) showed in a sample of 26 X-ray spectra of Seyfert~2 galaxies that 18 were best represented by an additional component ($kT = 0.1 - 1.5 \rm \, keV$), 4 objects required clumpy absorbers but no additional component, and 4 did not require any addition to the absorbed power law model.

It is remarkable that the type of model which fits the data best does not depend on the source type. There are type 1 and type 2 objects best modeled by an absorbing scenario (e.g. this work), and there are Seyfert 1 and Seyfert 2 galaxies for which a reflection model (e.g., Petrucci et al. 2007, Middleton et al. 2009, Wilkes, Pounds, \& Schmidt 2008), or an additional component like a cool corona (e.g., Dewangan et al. 2007, Guainazzi et al. 2005), represents the best model. This would indicate that the effects causing the soft excess do not depend on source type and inclination angle of the accretion disk. 
On the other hand, Pounds \& Wilkes (2007) pointed out that the effect of the soft excess is more pronounced in type 1 than in type 2 objects, and that a Seyfert 1.5 like 2MASS 0918+2117, shows indeed an intermediate excess. This would point to an origin of the soft excess far within the absorbing material responsible for the type 1 / type 2 distinction, and would support the hypothesis of Bianchi et al. (2009) that the observed X-ray spectra are influenced by several different absorbers. Larger well-defined samples, as presented e.g. in Guainazzi et al. (2005), with a consistent modeling approach are necessary though to verify this hypothesis.

\section{Summary and Conclusions}\label{conclusions}
We performed the first detailed X-ray analysis of the 3 Seyfert galaxies ESO 140-43, ESO~383-18, and UGC 3142 using {\it XMM-Newton} high resolution data in combination with {\it INTEGRAL}/IBIS and {\it Swift} BAT/XRT data.
The spectra of the three AGN presented indeed show what looks like a soft excess at energies E $< 2 \rm \,keV$, but which we have shown is due to complex absorption, without any additive emission to the power-law continuum.\newline
The 0.3--10 keV spectrum of the 3 Seyfert galaxies is well fit by a power law, absorbed by two partially covering neutral absorber for UGC~3142 and ESO~383-18, and by 3 clumpy ionized absorbers in the case of ESO~140-43. ESO~140-43 also shows evidence for a reflection hump.

The absorbing medium seen in ESO~383-18 could be a clumpy torus, plus a disc wind or Compton thin dust lanes. For UGC~3142 the observed absorbers might be part of a disc wind, or of the BLR. The variability of the flux and of the hardness ratio observed in the light curve of ESO~383-18 is not explainable by variations of the absorbers, and is likely related to changes of the continuum.

The absorbers of ESO~140-43 have different values of column densities and ionization states, according to the current picture of warm absorbers. 
Strong flux and spectral (accounted for by changes in the 3 warm absorbers) variability was detected between {\it ROSAT}, {\it EXOSAT}, and the two {\it XMM-Newton} observations of ESO~140-43. The spectral variability is likely due to the fact that the clumpy absorbers are moving, either as a part of a disk wind or of the BLR.
During the 2005-observation of ESO~140-43 variability of the covering factor of the LIA is detected on a time-scale of kilo-seconds, while on shorter time-scales the observed spectral variability cannot be explained by changes in the absorbers. Thus the effects of absorption on the variability are observable only on time-scales longer than few kilo-seconds, and are probably hidden by changes of the continuum on shorter time-scales.

}

We reported the first detection of the iron $K\alpha$ line in the spectra of ESO~383-18 and UGC~3142, and confirmed its presence in the spectrum of ESO~140-43. We tentatively associate an excess around 0.6 keV in the spectrum of ESO~383-18 with OVIII Ly$\alpha$ emission. More sensitive data will be necessary in order to confirm or rule out the presence of this line. The fact that we observe the iron $K\alpha$ lines, but not any additive sign of reflection (i.e. a reflection hump) for ESO~383-18 and UGC~3142, is probably due to the low significance of IBIS/ISGRI data and the hard X-rays variability, respectively.

In general there does not seem to be a common model explaining the presence of the soft X-ray excess in all the observed cases. Rather, it is likely that several components play a role, like reflection processes, complex absorption, and additional thermal emission. Not all of these components are observed in all objects and it remains to be clarified, whether the source type and inclination angle of the accretion disk has an influence on the type of soft excess observable in the X-ray spetra. It is anyway likely that some of the reported soft X-ray excesses in Seyfert galaxies are in fact caused by clumpy (neutral or ionized) absorbers, which allow a portion of the primary flux emitted by the central X-ray source to reach the observer directly. Thus the existence of the soft excess might be simply related, at least in some cases, to an incorrect absorption model, which over-estimates the absorbed flux below 2 keV.\newline
New broad band X-ray observations of these 3 Seyfert galaxies, e.g. by {\it Suzaku}, would be fundamental to probe the hypothesis that the absorption is responsible for the month scale variability, and to give better limits to the sizes and locations of the absorbers.

\begin{acknowledgements}
We would like to thank Simona Soldi, Chin-Shin Chang, Carla Baldovin and Marc T\"urler for their useful comments on the work, Andrew Taylor for reading and correcting the manuscript, and Tim Kallman for his help with XSTAR. This research has made use of the NASA/IPAC Extragalactic Database (NED) which is operated by the Jet Propulsion Laboratory, of data obtained from the High Energy Astrophysics Science Archive Research Center (HEASARC), provided by NASA's Goddard Space Flight Center, and of the SIMBAD Astronomical Database which is operated by the Centre de Donn\'ees astronomiques de Strasbourg. We would also like to thank the anonymous referees for their comments, which helped to improve this paper.
\end{acknowledgements}
References
\renewcommand{\baselinestretch}{1.0}
\newenvironment{references}{
    \begin{list}{}{\leftmargin 1.5em
                                           \itemindent -1.5em
                                           \itemsep 0pt
                                           \parsep 0.1cm
                                           \footnotesize
                                            }
                                           }{\end{list} }
\newcommand{\entry}{\item}


\begin{references}

%References

\entry
Abrassart, A. \& Czerny, B. 2000, A\&A 356, 475
\entry
Antonucci, R. R. J. 1993, ARA\&A, 31, 473
\entry
Armentrout, B. K., Kraemer, S. B., Turner, T. J. 2007, ApJ 665, 237
\entry
Arnaud, K. A. 1996, in: Astronomical Data Analysis Software and Systems V, eds. Jacoby G. and Barnes J., p17, ASP Conf. Series 101
\entry
Aretxaga, I., Joguet, B., Kunth, D., Melnick, J., \& Terlevich, R.~J.\ 1999, ApJ, 519, L123
\entry
Awaki, H., Koyama, K., Inoue, H., \& Halpern, J. P. 1991, PASJ, 43, 195
\entry
Bassani, L., Molina, M., Malizia, A., et al. 2006, ApJ 636, 65
\entry
Beckmann, V., Gehrels, N., Favre, P., et al. 2004, ApJ 614, 641
\entry
Beckmann, V., Petry, D., Weidenspointner, G. 2007, ATel 1264
\entry
Beckmann, V., Soldi, S., Ricci, C., et al. 2009, A\&A 505, 417
\entry
Bianchi, S., Piconcelli, E., Chiaberge, M., et al. 2009, ApJ 695, 781
\entry
Bianchi, S., \& Guainazzi, M. 2007, AIP Conference Proceedings, Volume 924, pp. 822-829
\entry
Blustin A. J., Page M. J., Fuerst S. V., et al. 2005, A\&A 431, 111
\entry
Blustin A. J., Kriss G. A., Holczer T., et al. 2007, A\&A 466, 107
\entry
Boller T., Meurs, E. I. A., Brinkmann, W., et al. 1992, A\&A 261, 57
\entry
Bottorff, M. C., \& Ferland, G. J., MNRAS 316, 103
\entry
Burrows, D. N., Romano, P., Falcone, A., et al. 2005, SSRv 120, 165
\entry
Caroli, E., Stephen, J. B., Di Cocco, G., Natalucci, L., Spizzichino, A. 1987, Space Sci. Rev., 45, 349
\entry
Chevallier, L., Collin, S., Dumont, A.-M., et al. 2006, A\&A, 449, 493
\entry
Cohen, R. D., Puetter, R. C., Rudy, R. J., et al. 1986, ApJ 311, 135
\entry
Courvoisier, T. J.-L., Walter, R., Beckmann, V., et al. 2003, A\&A, 411, 53
\entry
Courvoisier, T. J.-L. \& T$\ddot{\rm u}$rler, M. 2005, A\&A, 444, 417
\entry
Cusumano, G., La Parola, V., Segreto, A., et al. 2009, arXiv:0906.4788v1
\entry
Decarli, R., Dotti, M., Fontana, M., \& Haardt, F.\ 2008, MNRAS, 386, L15
\entry
Dewangan, G.~C., Griffiths, R.~E., Dasgupta, S., \& Rao, A.~R.\ 2007, ApJ, 671, 1284
\entry
Dickey, J. M. \& Lockman, F. G. 1990, ARA\&A 28, 215
\entry
Done, C.,  Mulchaey, J. S., Mushotzky, R. F. \& Arnaud, K. A. 1992, ApJ 395, 275
\entry
Done, C., \& Nayakshin, S.\ 2007, MNRAS, 377, L59  
\entry
Gallo, L. C.,  Tanaka, Y. \& Boller, T. 2004, MNRAS 353, 1064
\entry
Gehrels, N., Chincarini, G., Giommi, P., et al. 2004, ApJ 611, 1005
\entry
George, I. M., Turner, T. J., Mushotzky, R., et al. 1998, ApJ 503, 174
\entry
Ghosh, K. K. \& Soundararajaperumal, S. 1992, MNRAS 259, 175
\entry
Gierlinski, M., \& Done, C. 2004, MNRAS 349, L7)
\entry
Goldwurm, A., David, P., Foschini, L., et al. 2003, A\&A 411, L223
\entry
Grupe, D., Komossa, S., Gallo, L. C. 2007, ApJ 668, L111
\entry
Guainazzi, M., Matt, G., Antonelli, L. A., et al. 1998, MNRAS 298, 824 
\entry
Guainazzi, M., Matt, G., \& Perola, G. -C. 2005, A\&A 444, 119
\entry
Guainazzi, M., \& Bianchi, S. 2007, MNRAS 374, 1290
\entry
Guilbert, P. W. \& Rees, M.-J. 1988, MNRAS, 233, 475
\entry
Haardt, F. \& Maraschi, L. 1991, ApJ 380, L51
\entry
Haardt, F. \& Maraschi, L. 1993, ApJ 413, 507
\entry
Halpern, J.P. 1984, ApJ 281, 90
\entry
Holt, S. S., Mushotzky, R. F., Boldt, E. A., et al. 1980, ApJ 241, L13
\entry
Ishibashi, W., \& Courvoisier, T.~J.-L.\ 2009, A\&A 495, 113
\entry
Jansen, F., Lumb, D., Altieri, B., et al. 2001, A\&A 365, L1
\entry
Jim\'{e}nez-Bail\'{o}n, E., Guainazzi, M., Matt, G., et al. 2008, RMxAC 32, 131
\entry
Kallman, T. R., Liedahl, D., Osterheld, A., Goldstein, W., \& Kahn, S. 1996, ApJ 465, 994
\entry
Kallman, T. R., \& Bautista, M. 2001, ApJS 133, 221
\entry 
Kaspi S., Brandt W. N., George I. M., et al. 2002, ApJ 574, 643
\entry
Kinney, A.~L., Schmitt, H.~R., Clarke, C.~J., et al. 2000, ApJ 537, 152
\entry
Krivonos, R., Revnivtsev, M., Lutovinov, A., et al. 2007, A\&A 475, 775
\entry
Krolik, J. H., \& Begelman, M. C. 1986, ApJ 308, L55
\entry
Krongold, Y., Nicastro, F., Brickhouse, N. S., et al. 2003, ApJ 597, 832
\entry
Longinotti, A. L., Nucita, A. A., Santos, Lleo M., \& Guainazzi, M. 2008, A\&A, 484, L311
\entry
Longinotti, A. L., Costantini, E., Petrucci, P. O., et al. 2010, A\&A 510, 92
\entry
Magdziarz, P. \& Zdziarski, A. A. 1995, MNRAS 273, 873
\entry
Markowitz, A., Edelson, R., \& Vaughan, S. 2003, ApJ 598, 935
\entry
Marshall, F., Boldt, E., Halt, S., et al. 1979, ApJS 40, 657
\entry
Mason, K. O., Breeveld, A., Much, R., et al. 2001, A\&A 365, L36
\entry
Matt, G., Guainazzi, M., Perola, G. C., et al. 2001, A\&A 377, L31
\entry
Matt, G., Bianchi, S., De Rosa, A., et al. 2006, A\&A 445, 451
\entry
Matt, G., Bianchi, S., Awaki, H., et al. 2009, A\&A 496, 653 
\entry
McKernan, B., Yaqoob, T., \& Reynolds, C. S. 2007, MNRAS 379, 1359
\entry
Middleton, M., Done, C., Ward, M., Gierli{\'n}ski, M., \& Schurch, N.\ 2009, MNRAS, 394, 250
\entry
Mor, R., Netzer, H., Elitzur, M. 2009, ApJ 705, 298
\entry
Murray, N., Chiang, J., Grossman, S. A., \& Voit, G. M. 1995, ApJ 451, 498
\entry
Nandra, K., Fabian, A. C., George, I. M., et al. 1993, MNRAS 260, 540
\entry
Nandra, K., George, I. M., Mushotzky, R., Turner, T. J., \& Yaqoob, T. 1997, ApJ 476, 70
\entry
Netzer, H., Kaspi, S., Behar, E., et al. 2003, ApJ 599, 933
\entry
Nucita, A. A., Guainazzi, M., Longinotti, A. L., et al. 2010, arXiv:1003.1285
\entry
Padovani, P. \& Rafanelli, P. 1988, A\&A 205, 53
\entry
Pan, H. C., Stewart, G. C. \& Pounds K. A. 1990, MNRAS 242, 177
\entry
Pappa, A., Georgantopoulos, I., Stewart, G. C., \& Zezas, A. L. 2001, MNRAS, 326, 995
\entry
Petrucci, P.~O., Ponti, G., Matt, G., et al.\ 2007, The Multicolored Landscape of Compact Objects and Their Explosive Origins, AIPC, 924, 583 
\entry
Porquet, D., \& Dubau, J. 2000, A\&AS 143, 495
\entry
Pounds, K. A., Nandra, K., Fink, H. H. \& Makino, F. 1994, MNRAS 267, 193
\entry
Pounds, K. A., \& Wilkes, B.-J. 2007, \mnras 380, 1341
\entry
Piconcelli, E., Jim\'enez-Bail\'on, E., Guinazzi, M., et al. 2005, A\&A 432, 15
\entry
Puccetti, S., Fiore, F., Risaliti, G., et al. 2007, MNRAS 377, 607
\entry
Reeves, J. N., Nandra, K., George, I. M., et al. 2004, ApJ 602, 648
\entry
Reichert, G. A., Mushotzky, R. F., Petre, R., \& Holt, S. S., ApJ 296, 69
\entry
Reynolds, C.S. 1997, MNRAS 286, 513
\entry
Risaliti, G., Elvis, M., \& Nicastro, F. 2002, ASPC 258, 81
\entry
Risaliti, G., Elvis, M., Fabbiano, G., et al. 2006, ESASP 604, 655
\entry
Risaliti, G., Elvis, M., Fabbiano, G., et al. 2007, ApJ 659L, 111
\entry
Risaliti, G., Salvati, M., Elvis, M., et al. 2009, MNRAS 393, 1
\entry
Roming, P. W. A., Kennedy, T. E., Mason, K. O., et al. 2005, SSRv 120, 95
\entry
Saxton, R.~D., Turner,M.~J.~L., Williams, O.~R., et al. 1993, MNRAS, 262, 63
\entry
Schurch, N. J., \& Done, C. 2006, MNRAS 371, 81
\entry
Schurch, N.~J., \& Done, C.\ 2008, MNRAS, 386, L1 
\entry
Schurch, N.~J., Done, C., Proga, D. 2009, ApJ 694, 1
\entry
Soldi, S., Beckmann, V., Bassani, L., et al. 2005, A\&A 444, 431
\entry
Steenbrugge K. C., Kaastra J. S., Crenshaw D. M., et al. 2005, A\&A 434, 569
\entry
Str\"uder, L., Briel, U., Dennerl, K., et al. 2001, A\&A 365, L18
\entry
Titarchuk, L. 1994, ApJ, 434, 313
\entry
Tueller, J., Baumgartner, W. H., Markwardt, C. B., et al.\ 2010, ApJS accepted, arXiv:0903.3037
\entry
Turner, T. J., Nandra, K., George I. M., Fabian, A. C. \& Pounds, K. 1993, ApJ 419, 127
\entry
Turner, M. J. L., Abbey, A., Arnaud, M., et al. 2001, A\&A 365, L27
\entry
Turner, T. J., Reeves, J. N., Kraemer,  S. B. \& Miller, L. 2008, A\&A 483, 161
\entry
Turner, T.~J., \& Miller, L.\ 2009, \aapr, 17, 47
\entry
Vaughan, S., Edelson, R., Warwick, R. S., Uttley, P. 2003, MNRAS 345, 1271
\entry
Wilkes, B.-J., Pounds, K.-A., Schmidt, G.-D. 2008, ApJ, 680, 110


\end{references}
\end{document}